\documentclass[oupdraft, usenatbib]{bio}

\usepackage{caption}
\usepackage{lscape}
\usepackage{dirtytalk}
\usepackage{mathrsfs}
\usepackage{graphicx}
\usepackage{xr}
\usepackage{xcolor}
\usepackage{booktabs}
\usepackage{tabularx}
\usepackage{xparse}
\usepackage{longtable}

\allowdisplaybreaks

\usepackage{geometry}
\usepackage{lscape} 
\usepackage{setspace}
\usepackage{mathtools}

\usepackage{zwcommands,bm}
\usepackage{url,subfigure}

\usepackage{xr}

\begin{document}
\label{firstpage}

\title{\vspace{-2em}
Random Forest for Dynamic Risk Prediction of Recurrent Events: A Pseudo-Observation Approach}
\vspace{-1cm}\author{ABIGAIL LOE$^{\ast,1}$, SUSAN MURRAY$^{1}$, ZHENKE WU$^{1, 2}$\\[2pt]
\textit{$^1$Department of Biostatistics, University of Michigan, Ann Arbor, MI 48109, USA\\}
\textit{$^2$Michigan Institute for Data and AI in Society, University of Michigan, Ann Arbor, MI 48109, USA\\}
\vspace{-1em}{$^*$agloe@umich.edu}}

\markboth%
{A. Loe and others}
{Random Forest for Recurrent Events via Pseudo-Observation}

\maketitle
\footnotetext{To whom correspondence should be addressed.}
\vspace{-.5cm}
\begin{abstract}
{Recurrent events are common in clinical, healthcare, social and behavioral studies, yet methods for dynamic risk prediction of these events are limited. To overcome some long-standing challenges in analyzing censored recurrent event data, a 
recent regression analysis framework constructs a censored longitudinal data set consisting of times to the first recurrent event in multiple pre-specified follow-up windows of length $\tau$ \citep[][XMT]{Xia:StatMed:2019:regression}. Traditional regression models struggle with nonlinear and multi-way interactions, with success depending on the skill of the statistical programmer. With a staggering number of potential predictors being generated from genetic, -omic, and electronic health records sources, machine learning approaches such as the random forest regression are growing in popularity, as they can nonparametrically incorporate information from many predictors with nonlinear and multi-way interactions involved in prediction. 
In this paper, we (1) develop a random forest approach for dynamically predicting probabilities of remaining event-free during a subsequent $\tau$-duration follow-up period from a reconstructed censored longitudinal data set, (2) modify the XMT regression approach to predict these same probabilities, subject to the limitations that traditional regression models typically have, and (3) demonstrate how to incorporate patient-specific history of recurrent events for prediction in settings where this information may be partially missing.
We show the increased ability of our random forest algorithm for predicting the probability of remaining event-free over a $\tau$-duration follow-up window when compared to our modified XMT method for prediction in settings where association between predictors and recurrent event outcomes is complex in nature. We also show the importance of incorporating past recurrent event history in prediction algorithms when event times are correlated within a subject. The proposed random forest algorithm is demonstrated using recurrent exacerbation data from the trial of Azithromycin for the Prevention of Exacerbations of Chronic Obstructive Pulmonary Disease \citep{azithro}.}
{Censored data; Longitudinal data; Pseudo-observations; Random forest; Recurrent events}
\end{abstract}




%

\section{Introduction}
\label{s:intro}
Recurrent events are common in healthcare settings, such as exacerbations of pulmonary diseases, or bleeding events for chemotherapy patients. Clinicians are often interested in predicting the probability of remaining event-free until a patient's next check-up. Information from events over time helps improve precision of predictions relative to analyses of a single time-to-event. 
Prediction is also greatly enhanced by the large number of patient-specific predictors that are collected now, for instance, in the form of electronic health records data, gene expression levels, metabolomic data, etc. The very large dimensions of the predictors and the multivariate nature of recurrent event outcomes require careful thought for analysis, particularly when censoring of recurrent events is present.
A method that can utilize many potentially highly correlated risk factors of recurrent events for dynamic prediction would be of interest to modelers and practitioners.  

For dynamic prediction of an upcoming recurrent event, patient history of such events is often one of the most useful predictors (typically recurrent event-times within a patient are positively correlated).  A feature of recurrent event data is that each patient's event history at study time, $t$, becomes richer for making predictions as $t$ increases.  For instance in the Azithromycin for the Prevention of Exacerbation of COPD study \citep{azithro}, recurrent exacerbations were monitored over a year of follow-up; a cohort that we would like to use for making dynamic predictions of exacerbations. Unfortunately, patient history data for the most recent exacerbation time prior to study enrollment was not collected in this study. As a result,  this patient history data would not be available for dynamically predicting future events until later in study. How to include patient history data in dynamic predictions, when this data is only partially available at early follow-up times, has not been addressed in the recurrent-event setting.

The majority of literature in recurrent event analysis has been parametric or semi-parametric in nature, with specific modeled relationships between risk factors and recurrent event outcomes. Models have been designed for the analysis of recurrent events, gap times, and times-to-first-event across different follow-up windows, each with different assumptions imposed by the modeling paradigms \citep[e.g.,][]{ Andersen-Gill, Miloslavsky, MS_relapse, Mao_Lin_2016, recurrent_mods, kong_GEE, 3-levelSP, donglin-pi, jointmodeling, statistic-sinica, semiparametric_PropMeans, Xia:StatMed:2019:regression, aft_parametric_mod}. Such models have interpretable parameter estimates that can inform medical literature, generate hypotheses, and to some degree, predict outcomes. Disadvantages include limits to estimating these same parameters when there are many predictors or collinearity between important predictors that make unique solutions for corresponding parameter estimates difficult to obtain (often due to difficulty with inversion of nearly singular matrices; see, for example, \cite{HoerlKennard, GunstWebster}). In addition, methods that incorrectly assume independence of gap times between recurrent events measured on an individual give biased results due to induced dependent censoring. 

A few authors propose pseudo-observation approaches to parametrically or semi-parametrically model censored recurrent event data \citep{DPO_Paper, Xia:StatMed:2019:regression}, replacing censored outcomes with tailor-made pseudo-observations suitable for obtaining desired parameter estimates from models available for uncensored data. \cite{DPO_Paper} developed pseudo-observations for modeling probabilities that the $k$-{th} recurrent event occurred in a particular follow-up window. \cite{Xia:StatMed:2019:regression} developed pseudo-observations for modeling the restricted mean event-free time in a particular follow-up window. Both of these approaches are versions of landmark analysis, where outcomes for analysis are measured relative to landmark times.  \cite{Xia:StatMed:2019:regression}'s landmarking approach transforms censored recurrent event data into a censored longitudinal data structure with restricted mean pseudo-observations defined over a series of follow-up windows that can be analyzed using standard longitudinal data analysis methods. 

As to the goal of dynamically predicting recurrent event risk based on traditional censored recurrent event data with many potentially colinear predictors, perhaps with non-linear associations with recurrent event outcomes, we are drawn to the nonparametric nature of random forest algorithms. Machine learning methods, and in particular random forest algorithms, have gained widespread acceptance as a method of building predictions based on complex relationships of biomarkers (first proposed by \cite{CanonicalCART}, applied more recently in statistical literature such as 
\citet{zhao2020incorporating}, \citet{JASArf1}, 
and reviewed in \citet{JoesPaper}). These algorithms naturally accommodate non-canonical relationships between covariates and outcomes, an attractive feature in an age of increasingly high-dimensional data where a traditional modeler cannot realistically find all the interactions, higher-order covariates and non-linear patterns that may be predictive of an outcome. By using information across different sets of covariates within each of the generated trees of a random forest, a modeler can side-step several traditional pitfalls of parametric and semi-parametric models: (1) multicolinearity of predictors and (2) limited ability to include a large number of covariates. In particular, \citet{zhao2020incorporating} showed that for a single event time, imposing a censored longitudinal data structure introduced by \citet{TayobMurray:Biostat:2015} and incorporating ideas from pseudo-observation literature, random forest regression can address nonlinear relationships  between covariates and outcomes for dynamic prediction of event-free periods. Machine learning methods for censored recurrent event data are still in their infancy; we are not aware of any machine learning methods available for the analysis of recurrent event data.  

In our manuscript we incorporate ideas from \citet{Xia:StatMed:2019:regression} on building a censored longitudinal structure for censored recurrent event data using pseudo-observations.  Our methods make several important advances from this point: (1) To better reflect the goals of dynamic prediction, we develop pseudo-observations for analysis of event-free probabilities over $\tau$-duration follow-up windows measured from regularly spaced landmark times over the duration of the entire follow-up period. (2) We develop a modified version of the Xia, Murray and Tayob (XMT) model where the estimand of interest is the probability of being event-free over the subsequent $\tau$-duration window, rather than a $\tau$-restricted mean. (3) We propose and evaluate different methods for including historical event rate information in these analyses with a multiple imputation approach for addressing missing event history data prior to the initiation of follow-up, and finally, (4), we describe a novel random forest algorithm with increased prediction accuracy to perform dynamic prediction of event-free probabilities over $\tau$-duration follow-up windows applicable to censored recurrent event data using elements from advances (1) and (3). This final random forest tool addresses many difficult issues facing analysts of recurrent event data mentioned above and we compare the performance of dynamic prediction using our modified-XMT model versus our random forest algorithm. 

The rest of this paper is organized as follows. In Section \ref{s:model}, we describe how to transform censored recurrent event data to a censored longitudinal data framework for analysis and offer examples of useful patient history covariates.  In Section \ref{s:POs}, we describe how to construct pseudo-observations based on the censored longitudinal data structure given in Section \ref{s:model}. Using these pseudo-observations, we describe dynamic prediction based on our modified version of the Xia, Murray and Tayob (XMT) model in Section \ref{s:modifiedXMT}. In Section \ref{s:HRF}, again using pseudo-observations defined in Section \ref{s:POs}, we describe how to perform dynamic prediction via a random forest algorithm applicable to longitudinal data (i.e., the constructed pseudo-observations). Model evaluation metrics are summarized in Section \ref{s:evaluation}. In Section \ref{s:SimStudy}, we use simulations to compare our modified XMT regression approach and our proposed random forest approach for the analysis of censored recurrent event data, and evaluate different methods for including historical event rate information in these analyses. We apply both the modified XMT and our recurrent event random forest algorithm to data from the study of Azithromycin for the Prevention of Exacerbation of COPD study in Section \ref{s:data-app}. We conclude with a discussion in Section \ref{s:discussion}.

\vspace{-1cm}
\section{Notation and Censored Longitudinal Data Structure}
\label{s:model}
For individual $i$, denote recurrent event times $T_{i, 1}< T_{i, 2}<\ldots$, that are potentially censored at time $C_i$, where $C_i$ is independent of $T_{i,j}$, $i = 1, ..., n$, and $j \geq 1$. 
Independent censoring mechanisms are common, for instance, in settings where participants are identified during a finite accrual period and followed for a predetermined period (often based on research funding). 
Observed data for each individual is $X^*_{i, j}= \min(T_{i, j}, C_i)$, $i = 1, \ldots, n$, and $j \geq 1$. These event times are sometimes converted to gap times between recurrent events $G_{i,j}= T_{i, j+1}-T_{i , j}$, $i = 1, \ldots, n$, $j \geq 1$ for analyses. However, a well-known challenge is the dependence between gap time random variables, $G_{i, j}$, and corresponding censoring times, $C_i-T_{i, j-1}$ \citep{dependentGapTime}. 


One approach to avoiding dependent censoring in analyses of recurrent events was proposed by \citet{TayobMurray:Biostat:2015}, who transformed correlated recurrent event times into a censored longitudinal data structure. First, a series of follow-up window start times are pre-specified to $\mathcal{T}=\{t_0, t_1, t_2, \ldots\}$, with $t_j - t_{j-1} = a$, $j \geq 1$; the start times can be as close as $a=1$ day apart, if desired. Computational speed of the analysis is usually the limiting factor in choosing $a$. \citet{XiaMurray:Biostat:2019:commentary} found that $a$ equal to one-third of the mean gap time in the control group resulted in capturing 90\% of the recurrent events from the original data set in at least one follow-up window. In practice, more frequent follow-up window start times have the potential for more precise estimation, but with diminishing returns once all recurrent events have been captured in at least one follow up window.

In this paper, at each of the pre-specified ``check-in time points'' in $\mathcal{T}$, we will focus on estimating the probability of remaining recurrent-event-free over a subsequent follow-up window of length $\tau$, where $\tau$ is user-specified and ideally informed by clinical expertise. Although the check-in time points are shared across subjects by design, potential censoring may result in different numbers of effective check-ins; we therefore denote the subject-specific sets of check-in time points by $\{\mathcal{T}_i\}_{i=1}^{n}$, where $\mathcal{T}_i = \{t_0, t_1, ... t_{b_i}\}$. In the case of no censoring, we chose $t_{b_i}$ so that the final follow-up interval from $t_{b_i}$ to $(t_{b_i}+\tau)$ does not exceed the study period; in the case with censoring, we choose $t_{b_i}< C_i$.
We follow \citet{TayobMurray:Biostat:2015} and construct the censored longitudinal data that is essential for creating pseudo-observations in Section \ref{s:POs}. We repeat the following two steps at all the check-in time points $t\in \mathcal{T}_i$, $i=1, \ldots, n$:

\vspace{-0.25cm}

\begin{itemize}
\item[i)] Identify the first recurrent event occurred after $t$; let $\eta_i(t)= \min \{j: X^*_{i,j}\geq t, j = 1, \ldots, n_i\}$ be its index among all the observed recurrent events for subject $i$. Define the residual recurrent-event-free time by $T_i(t) = T_{i, \eta_i(t)} - t$ with the corresponding censoring random variable $C_i(t) = C_i - t$, which importantly remains independent of $T_{i}(t)$;

\item[ii)] Construct the observed censored longitudinal data by $\{X_i(t) =X^*_{i, \eta_i(t)}-t, \delta_{i}(t) = I\{X_i(t)\leq C_i(t)\}: i = 1, 2, \ldots, n\}$.
\end{itemize}

\vspace{-0.25cm}

\begin{figure}
\centering
\includegraphics[width=\textwidth]{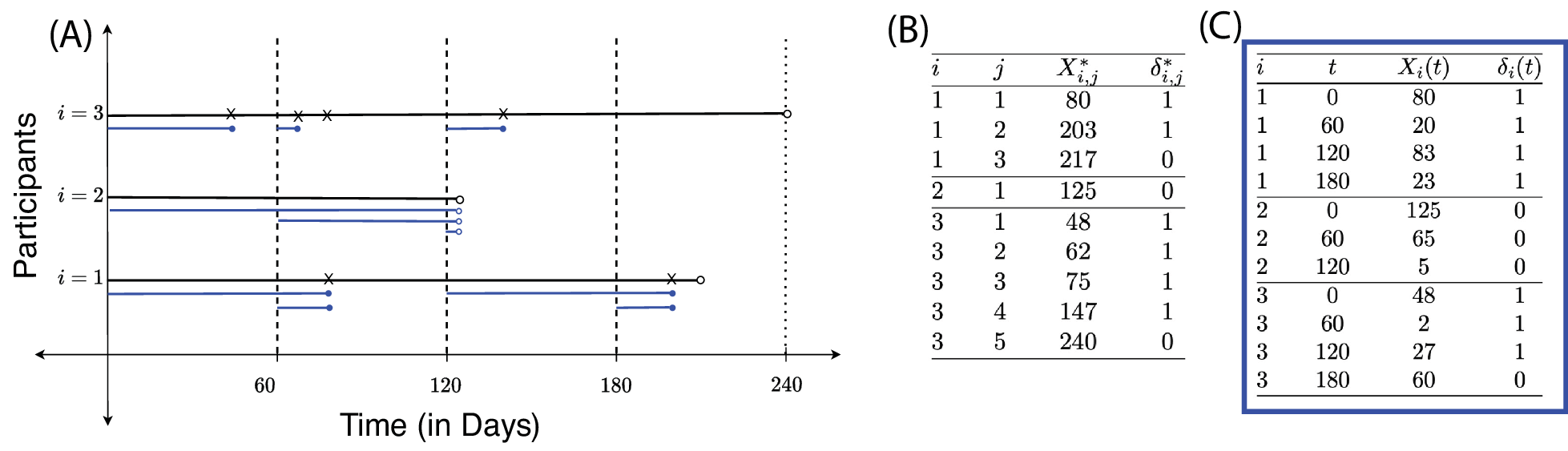}
\caption{Panel (A) displays a visualization of recurrent events in both traditional format (black) and censored longitudinal format (blue). Start times of follow-up windows ($\mathcal{T} = \{0, 60, 120, 180 \text{ days}\}$, with $a = 60$; this step does not concern $\tau$) are highlighted by vertical dashed lines. Traditional observed recurrent event data for individuals $i = 1, 2, 3$ are observed in Panel (B), while the observed censored longitudinal data for these individuals is displayed in Panel (C). All participants were administratively censored after 8 months (240 days) of follow-up. Potentially censored event times are denoted by the random variable $X^*_{i,j}$ in Panel (B), with the corresponding censored longitudinal measures denoted by $X_i(t)$ in Panel (C). Censoring indicators are denoted by  $\delta^*_{i,j}$ or $\delta_i(t)$ in Panels (B) and (C) respectively. In this visualization, Participant 2 has no recorded events before censoring time, denoted $C_2$. Additionally, while $X^*_{3,3}$ is technically observed, it is not included in the censored longitudinal data set because it is the second observation in the window, and there is no $t$ such that $X_i(t) = X^*_{3,3}-t$. Besides potential correlation between separate event times seen in the same individual, this data structure induces correlation due to the same event contributing to the final censored longitudinal values in more than one follow-up window. For example, $X^*_{1,2}$ produces two measures in the longitudinal data structure, $X_1(120) = 83$ and $X_1(180) = 23$.}\label{fig:data_viz}
\end{figure}

 Figure \ref{fig:data_viz} uses three hypothetical subjects to illustrate the above two steps of converting traditional recurrent event data $\{X^*_{i, j}, \delta_{i, j}\}$ into the censored longitudinal data format $\{X_i(t), \delta_i(t): t\in \mathcal{T}_i\}$. 
 Our prediction target can now be formally expressed as $P(T_i(t) \geq \tau|\bZ_i(t))$, i.e., at a pre-specified time point $t$, the probability of remaining  recurrent-event-free over a subsequent follow-up window of length $\tau$ given $\bZ_i(t)$, a set of time-invariant or time-varying predictors observed up to time $t$. In the following, we will occasionally suppress the subscript $i$ and write $P(T(t) \geq \tau | \bZ(t))$ for ease of presentation. 

\vspace{-.5cm}
  \subsection{Constructing Historical Recurrent Event Information as Predictors}
 \label{sec:construct_history}
 
 An important time-dependent covariate available in longitudinal studies is past recurrent event history, particularly when recurrent events within a subject are highly correlated. However, it is not always the case that history covariates are fully known at the beginning of study follow-up. Thus we consider two different types of history covariates (1) \textit{full} history covariates, where $\bZ_i(t)$ is known for each follow-up window starting at $t_0 \in \mathcal{T}_i$ or (2) \textit{partial} history covariates where $\bZ_i(t)$ is unavailable for analysis at $t_0$, but recorded for all $t> t_0$. 
 In setting (2), we perform a multiple imputation analysis where, for each imputed dataset, $m=1,\dots,M$,  random variables of interest prior to $t_0$ for individual $i$ are independently sampled from an appropriate distribution, controlled by parameters that are themselves sampled from uniform distributions consistent with clinician knowledge of recurrent event behavior in the population.
 
 For each follow-up window starting at $t \in \mathcal{T}_i$, we define predictors based on the history recurrent events which often can be predictive of future events. Three definitions are presented below based on the motivating application in this paper, though others are possible. 
 
 First, consider a dynamic predictor that averages the observed $\tau^*$-restricted event times at the check-in time points prior to time $t$. Specifically, in the history before $t$, identify those check-in time points: $\mathcal{T}^*_i(t)=\{t^*\in \mathcal{T}_i\cup \mathcal{T}^0_i: t^* + \tau^* \le t\}$, where $\tau^*$ may differ from the original $\tau$. Here $\mathcal{T}^0_i$ is a set of check-in time points before time $t_0$ that are evenly spaced with $a^*$ time units away; it is introduced here solely for constructing predictors based on history information prior to time $t_0$. Here $a^*$ may differ from $a$ used in the construction of the censored longitudinal data structure at the beginning of Section \ref{s:model}.  The $\tau^*$-restricted event time is then $\min\{ T_i(t^*), \tau^*\}$. We average them over all the $t^*$ in $\mathcal{T}_i^*(t)$ to construct:
 \begin{equation} \label{eq:fullZ}
 H_{f,i}^{(1)}(t)= \left.{\sum_{t^* \in \mathcal{T}^*_i(t)} \min\{ T_i(t^*), \tau^*\}}\right/{|\mathcal{T}^*_i(t)|}.
 \end{equation}
 
 

Note that in Equation (\ref{eq:fullZ}), at $t=t_0$, the definition assumes there exists historical recurrent event information for analysis. Otherwise, patient history prior to $t_0$ can be supplemented via (multiple) imputations of recurrent event times, $\Tilde{T}_{i, j}$, which are then converted to imputes of the censored longitudinal random variable, denoted as $\tilde{T}_i(t^*)=\Tilde{T}_{i, j}-t^*$ for $t^*\in \mathcal{T}^*_i(t_0)$. We now can define a partially observed version of Equation (\ref{eq:fullZ}):
\begin{equation} \label{eq:partialZ}
 H_{p,i}^{(1)}(t)=\left.{\sum_{t^* \in \mathcal{T}^*_i(t)} 
 I(t^* \le  t_0) \times \min\{ \Tilde{T}_i(t^*), \tau^*\}+ I(t^*>t_0) \times \min\{ T_i(t^*), \tau^*\} }\right/{|\mathcal{T}^*_i(t)|}.
 \end{equation} 

 Second, at check-in time $t$, we may construct a predictor based on how long in the past $\tau^*$ period did the most recent event occur, if it exists. With the full history recurrent event information available, the time since the most recent exacerbation at $t$ is 
    \begin{equation*}
       H^\dag_{f, i}(t) = \begin{cases}
           \min\limits_{j: T_{i,j}< t}\{ | t - T_{i, j}|\} & \text{if } \{j : T_{i, j}< t\} \neq \emptyset;\\
           \tau^* & \text{if } \{j : T_{i, j}< t\} = \emptyset.
       \end{cases}
    \end{equation*} This information can be coded as a continuous history variable, $\min \{\tau^*, H^\dag _{f,i}(t)\}$, or be coded as a categorical variable, for instance:
\begin{align} 
    H_{f, i}^{(2)}(t) = &\ 4 \times I\left\{ 0 \mbox{ days} \le H^\dag_{f, i}(t) < 31 \mbox{ days}\right\}+
    3 \times I\left\{ 31\mbox{ days} \le H^\dag_{f, i}(t) < 93 \mbox{ days}\right\} + \notag \\ 
    &\ 2 \times I\left\{ 93 \mbox{ days} ^* \le H^\dag_{f, i}(t) < 182\mbox{ days}\right\} + I\left\{ 182 \mbox{ days} \le H^\dag_{f, i}(t) < \tau ^*\right\},\label{eq:fullTimeSince}
\end{align}
where increasing levels of $H_{f, i}^{(2)}(t)$ indicate shorter periods without recurrent events in the past, and $H_{f, i}^{(2)}(t) = 0$ when $H^\dag_{f, i}(t)\ge \tau^*$.
When the full history of recurrent information is not available for analysis at $t_0$, one may impute $\Tilde{T}_{i,j}$ and identify the most recent imputed event time before $t_0$ as $\max \Tilde{T}_i$. We then define a partial history version of Equation (\ref{eq:fullTimeSince}) as:
\begin{equation*}
    H^\dag_{p, i}(t) = \begin{cases}
        |t- \max\Tilde{T}_i| & \text{if } t \leq t_0; \\
        |t-\max\Tilde{T}_i|  & \text{if } t > t_0 \text{ and } \{j: T_{i, j}< t\}= \emptyset; ~\textrm{and}~ \\
        \min\limits_{j: T_{i, j}< t} \{|t-T_{i, j}|\} & \text{if } t> t_0\text{ and, } \{j : T_{i, j}< t\} \neq \emptyset.\\
    \end{cases}
\end{equation*}
Again, similar to the $H^{(2)}_{f,i}(t)$ case, we may code this variable as the continuous version $H_{p,i}^{(2)}(t) = \min\{\tau^* , H^\dag_{p,i}(t)\}$ or use a categorical form, as in Equation (\ref{eq:fullTimeSince}), replacing $H^\dag_{f,i}(t)$ with $H^\dag_{p,i}(t)$.

Third, the history recurrent event information predictor can be based on the frequency with which a recurrent event occurred within $\tau^*$ time units at each of the past check-in time points $t^*\in \mathcal{T}^*_i(t)$. If all the historical recurrent event information is available for analysis, we define 
 \begin{equation} \label{eq:fullCldfRate}
 H^{(3)}_{f,i}(t)=\left.{\sum_{t^* \in \mathcal{T}^*_i(t)} I \{{T}_i(t^*) \le \tau^* \} }\right/{|\mathcal{T}^*_i(t)|}.
 \end{equation} 
 In the case where events prior to time $t_0$ are imputed as described above, we define
 \begin{equation}\label{eq:partialCldfRate}
H^{(3)}_{p, i}(t)=\left.{\sum_{t^* \in \mathcal{T}^*_i(t)} [
 I(t^*  \le  t_0) \times I \{\tilde{T}_i(t^*)< \tau^*\} + I(t^*>t_0) \times I\{T_i(t^*)< \tau^*\} ]}\right/{|\mathcal{T}^*_i(t)|}.
 \end{equation}

Relatedly, analysis using multiply-imputed partial history variables included in the set of dynamic predictors $\bZ_i(t)$ requires combination via Rubin's rule, where appropriate. Briefly, Rubin's rule allows an analyst to combine estimates of a parameter of interest, say $\nu$, from separate analyses of imputed datasets, summarizing results with an overall point estimate and corresponding variance estimate \citep{RubinsRulesCite}. Let $m = 1,2, \ldots, M$ denote the number of complete (potentially imputed) datasets  with point estimates $\hat{\nu}_m$, and estimated variance $\hat{\sigma}_m^2$. The combined point estimate is $\Bar{\nu}= \frac{1}{M}\sum_{m=1}^M \hat{\nu}_m$. The pooled variance, $V_t$, associated with $\Bar{\nu}$ has two components: the within- and the between-imputation variance. We define $V_t = \frac{1}{M}\sum_{m = 1}^M \hat{\sigma}_m^2 + \frac{(M+1)}{M(M-1)}\sum_{m = 1}^M (\hat{\nu}_m-\Bar{\nu})^2$. Relying on theory of asymptotic tests for large $n$, $\frac{\Bar{\nu}^2}{V_t}\sim \chi^2_{(1)}$. For more details, see \citet{rubin1996multiple}.

\vspace{-.8cm}
\section{Pseudo-Observations for Dynamic Prediction}
\label{s:POs}
 In this section, we (1) briefly review the heuristics behind the use of pseudo-observations in censored data settings and (2) develop pseudo-observations appropriate for dynamic prediction based on the censored longitudinal data structure described in Section \ref{s:model}.

Pseudo-observations are a variation of jack-knife methodology applicable to censored time-to-event outcomes \citep{andersen2003generalised}. This technical device has been used successfully for a single time-to-event and recurrent events in manuscripts such as: \citet{Andrei_Murray_2007, recurrent_mods, Xia:StatMed:2019:regression, zhao2020incorporating}. Essentially a pseudo-observation of a censored outcome, once created, can be used in regression contexts as if it was an uncensored outcome. 



The theoretical justification behind the use of pseudo-observations typically centers on an estimand of interest for each individual $i$, which in our dynamic prediction setting is $P\{T_i(t) \ge \tau | \bZ_i(t)\},$ i.e., the conditional probability of being event free in the follow-up window of length $\tau$ starting at $t$ given an individual's covariates, $\bZ_i(t),$ known at the beginning of this follow-up window for $i=1,\ldots,n(t)$, where $n(t)$ is the number of individuals at risk at time $t$. This conditional probability can be equivalently expressed as a conditional expectation, $E[I \{T_i(t) \ge \tau\} | \bZ_i(t) ]$, where $I \{T_i(t) \ge \tau\}$ given $\bZ_i(t)$ is a Bernoulli$[P\{T_i(t) \ge \tau | \bZ_i(t)\}]$ random variable that is not completely observed if $T_i(t)$ is censored prior to $\tau$, and $E[I \{T_i(t) \ge \tau\} | \bZ_i(t) ]$ is a random variable that depends on $\bZ_i(t)$ and its corresponding probability distribution.  The law of total expectation states that the mean of $E[I \{T_i(t) \ge \tau\} | \bZ_i(t) ]$ taken across the distribution of $\bZ(t)$ is algebraically equivalent to $E[I \{T(t) \ge \tau \}]=P\{T(t) \ge \tau\}$, which is the (unconditional, marginal) expected value of $I \{T(t) \ge \tau\}$ in the absence of knowledge of $\bZ(t)$. That is, \begin{equation}
    \label{keyPoEq2}
      P\{T(t) \ge \tau\} =  E[P\{T(t) \ge \tau |\bZ(t)\}],
\end{equation}
where the outermost expectation on the right hand side of this expression is taken with respect to the population distribution of $\bZ(t)$.  

The intuition behind creating a pseudo-observation for individual $i$ comes from applying Equation (\ref{keyPoEq2}) to two different empirically observed distributions of $\bZ(t)$ based on the cohort of $n(t)$ individuals contributing data during the follow-up window starting at $t$: (1) the empirically observed distribution of $\bZ(t)$ based on the entire cohort, $j=1,\ldots,n(t),$  and (2) the empirically observed distribution of $\bZ(t)$ based on this cohort, excluding individual $i$. For cohort (1), Equation(\ref{keyPoEq2}) becomes
\begin{equation}\label{keyPoEq3}
      P\{T(t) \ge \tau\} =  \frac{1}{n(t)} \sum_{j=1}^{n(t)} P\{T_j(t) \ge \tau |\bZ_j(t)\}.
\end{equation}
For cohort (2), Equation(\ref{keyPoEq2}) becomes
\begin{equation}\label{keyPoEq4}
      P^{(-i)}\{T(t) \ge \tau\} =  \frac{1}{n(t)-1} \sum_{j \neq i}^{n(t)} P\{T_j(t) \ge \tau |\bZ_j(t)\},
\end{equation}
where we use the notation $P^{(-i)}\{T(t) \ge \tau\}$ to indicate that individual $i$ being excluded from the empirical distribution of $\bZ(t)$ affects quantities on both sides of this expression. Taking Equations (\ref{keyPoEq3}) and (\ref{keyPoEq4}), together and solving for $P\{T_i(t) \ge \tau | \bZ_i(t)\}$, we get 
\begin{equation}\label{keyPoEq45}
P\{T_i(t) \ge \tau | \bZ_i(t)\} = n(t)P\{T(t) \geq \tau \}- \{n(t)-1\} P^{(-i)}\{T(t) \geq \tau\},
\end{equation}
where the left hand side of this expression is the dynamic prediction probability for patient $i$ that we would like to obtain and the right-hand side involves marginal population quantities.  The right-hand side of this Equation can be estimated completely nonparametrically using Kaplan-Meier estimates $\hat{P}(T(t) \geq \tau)$ for $P(T(t) \geq \tau )$, and $\hat{P}^{(-i)}(T(t) \geq \tau)$ for $P^{(-i)}(T(t) \geq \tau)$; the formulae are provided in Supplemental Materials Section A1. An appropriate pseudo-observation for dynamic prediction in the recurrent event setting is one that has approximately the same mean as $I [T_i(t) \ge \tau]$ given $\bZ_i(t)$, the uncensored random variable of interest.
The pseudo-observation,
\begin{equation}\hat{S}^\tau_{i}(t) = n(t)\hat{P}(T(t) \geq \tau )- [n(t)-1] \hat{P}^{(-i)}(T(t) \geq \tau ),
\end{equation}
has asymptotic mean, $P[T_i(t) \ge \tau | \bZ_i(t)],$ which meets this criterion. In fact, in the special case with no censoring, all pseudo-observations, $\hat{S}^\tau_{i}(t),$ reduce algebraically to $I(T_i(t) \ge \tau), i=1,\ldots,n(t).$
\vspace{-.5cm}
\section{Semiparametric Model for Dynamic Prediction (Modified XMT Model)}
\label{s:modifiedXMT}
Pseudo-observations, $\hat{S}^\tau_{i}(t), t \in \mathcal{T}_i,i=1,\ldots,n(t)$, may be used as observed outcomes in a variety of analyses typically reserved for uncensored data, including Generalized Estimating Equation (GEE) models and longitudinal random forest algorithms. 
Consider the marginal model (estimated by GEE), 
\begin{equation}\text{logit} [\hat{S}_i^\tau(t)] = \beta^\top \bZ_i(t) \label{mod-XMT},
\end{equation} where outcomes for individual $i$, $\{ \hat{S}_i^\tau(t), \mbox{ for } t \in \mathcal{T}_i \}$ are assumed to have an unstructured correlation matrix.  Because the originally published XMT model used pseudo-observations for $\tau$-restricted means and not our custom made pseudo-observations for dynamically predicting recurrent event probabilities, $\hat{S}^\tau_{i}(t), i=1,\ldots,n(t)$, we call model (\ref{mod-XMT}) ``the modified XMT (m-XMT) model'' in this manuscript.
Existing software can be used to fit this model. Estimated dynamic prediction probabilities from this model are $\hat{P}[T_i(t) \ge \tau | \bZ_i(t)]=\textrm{logit}^{-1}\{\hat{\beta}^\top \bZ_i(t)\}$.
This proposed regression model is a semiparametric model for dynamic prediction of recurrent event probabilities, subject to the same limitations described for parametric and semiparametric models in the introduction; that is, difficulty in the regression specification when there are many possibly highly correlated predictors.
\vspace{-1cm}
\section{Random Forest for Dynamic Prediction (RFRE.PO algorithm) }
\label{s:HRF}
 
Regression trees are a popular approach to incorporating multi-way interactions among predictors by finding groups of observations that are similar. A tree is grown in a few steps where at each step a new branch sorts the data leftover from the preceding step into bins based on one of the predictors. The sequential branching slices the space of predictors into rectangular partitions and approximates the true outcome-predictor relationship with the average outcome within each partition. Therefore, to grow a tree is to find bins that best discriminate among the outcomes. The specific predictor and the value split at each branching are chosen to minimize prediction error. 

The number of possible trees is combinatorially large, challenging efficient global optimization. Greedy algorithms have been developed to approximate the optimal global tree by myopically optimizing prediction error at the start of each branch. We will focus on binary trees in this paper for their popularity and effectiveness. The loss associated with the prediction error for a branch is often termed ``impurity'' which measures how similarly the observations behave on either side of the binary split. The branching procedure halts, for example, when the number of observations in a terminal node or the number of terminal nodes reaches respective thresholds. Advantages of a regression tree include invariance to monotonic transformation of predictors, flexible approximation to potentially severe nonlinearities, and the capacity to approximate $L-1$ way interactions for a tree with depth $L$. 

To overcome potential overfitting, ensemble methods can be used to combine predictions from many trees into a single prediction. Random forest is such a ensemble regularizer based on Breiman's bootstrap aggregation \citep{breiman2001random}, or ``bagging,'' which averages over multiple predictions obtained from trees grown from multiple bootstrap samples hence stabilizing the overall prediction. Random forest is a variation of bagging designed to further reduce the correlation among trees grown using different bootstrap samples by ``dropout'' which considers only a randomly drawn subset of predictors for splitting at each potential branch. Such a strategy ensures that early branches for some trees will not always split on predictors that offer the most gain in prediction accuracy. This reduces the correlation among predictions by multiple trees to further improve the variance reduction relative to the standard bagging along with reduced computational cost.

Among the many variants of random forest, we are particularly interested in random forests for longitudinal data, as recurrent event analysis uses data obtained from subjects over time. \citet{hrfRef} proposed an algorithm for tree-based ensemble methods that takes into consideration the dependence structure inherent in longitudinal data analysis. We provide a primer on this specific variation of random forest.

\begin{figure}
    \centering \includegraphics[width=\textwidth]{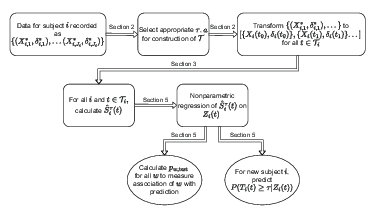}
    \caption{A flowchart depicting the analysis pipeline of \texttt{RFRE.PO}. Labels above arrows correspond to the section which describes the corresponding step, as well as notation, in greater detail. An analyst begins with data recorded as recurrent event times for subject $i$, $j=1, 2, \ldots, J_i$ before constructing an appropriate $\mathcal{T}$, and transforming data into the censored longitudinal framework. Upon creating such a structure, the analyst 
    calculates pseudo-observations, which are ultimately fed to a historical random forest. Recall that $p_{w, \text{test}}$ refers to the permutation test statistic described in Section \ref{s:HRF}, and $w$ refers to a variable selected for permutation} \label{fig:flowchart}
\end{figure}

Let $[\hat{S}_{i}^\tau(t)$, $\bZ_{i}(t)]$ be the data for individual, $i$, pertaining to the $\tau$-duration follow-up window starting at $t$, where $\hat{S}_{i}^\tau(t)$ is a pseudo-observation as defined in Section \ref{s:POs} and $\bZ_{i}(t)$ is a set of $p$-dimensional covariates available at time $t \in \mathcal{T}_i$ for $i = 1,2, ..., n$; here we use $\cT_i$'s introduced in Section \ref{s:model} with shared measurement timings to align more closely with our construction in Section \ref{s:POs}, although the algorithm presented herein is applicable to scenarios with subject-specific measurement timings. When pseudo-observations of probability of time-to-first-recurrent-events are applied to random forests for prediction of the structure described here, we call the algorithm ``\texttt{RFRE.PO}'' (short for ``Random Forest for Recurrent Events based on Pseudo-Observations''); we depict the analysis pipeline in  Figure \ref{fig:flowchart}.  
Also note that the $\bZ_{i}(t)$ may include time $t$. The following algorithm assumes $\hat{S}_i^\tau(t) = g\{\bZ_i(t)\}+\epsilon_i(t)$ with $\EE[\epsilon_i(t)]=0$ and is designed to estimate $g(\cdot)$ which describes a potentially complex relationship between the outcome and the predictors. 

For $b = 1, \ldots, B$, we repeat Step 1 and 2 below ($B=500$ is used in this manuscript):

\begin{itemize}
\item[Step 1] (Bootstrap): Generate a bootstrap sample $\{(\hat{S}_{i}^\tau(t), \bZ_i(t)), (i,t) \in Bootstrap(b)\}$ from the original data set, where $Bootstrap(b)$ is the $b$-th bootstrap sample; $Bootstrap(b)$ is obtained by first sampling $n$ subjects with replacement, followed by selecting all check-in time points $t \in \{t_0, t_1, ..., t_{b}\}$ for each bootstrapped subject $i$ \citep{hrfRef}. The data corresponding to individuals in $Bootstrap(b)$ is called \say{bagged}, and data not in the sample is called \say{out-of-bag.} 

\item[Step 2] (Grow a tree): Initialize the tree stump by grouping all observations together. 
\begin{itemize}
\item[2a.] (Form split variables) At a potential branching, randomly select $m$ predictors from the $p$ predictors with $m\ll p$; in this paper we use the default choice of $m=\sqrt{p}$ which works well in our simulation and data analysis. Let the randomly selected predictors be $\{Z_{i,s_1}(t), \ldots, Z_{i,s_m}(t)\}$ at a particular potential branching for bagged observation $(i,t)$ where $(s_1,\ldots, s_m)$ are the indices for the subset of $m$ predictors. For ease of presentation, we denote it by $\bW=(W_1, \ldots, W_m)$. For a continuous or categorical observed covariate $W_k$, construct a grid of all observed values of the covariate, $\Tilde{W}_k = \{w_{kl}, l = 1, \ldots, L_k \}$. Consider all possible thresholds in $\Tilde{W}_k$ and their corresponding indicators, $\{I(Z_{i,s_k}(t)< w_{kl}), (i,t) \in Bootstrap(b), w_{kl} \in \Tilde{W_k}\}$ for subject $i$ at check-in time $t$ in the bootstrap sample $Boostrap(b)$. These derived indicator functions for continuous and categorical covariates, along with covariates in $\bW$ that were already binary, are \say{potential split variables}. 

\item[2b.] (Choose branching): Choosing a covariate among $\bW$ and the split value $w_{kl}$ will result in two daughter nodes, $\{N_{w_{kl}, 1}, N_{w_{kl}, 0}\}$. Criteria for selecting the best variable and split vary by outcome type and metric of prediction accuracy.
For continuous outcomes, the split with minimal sum of squared errors (SSE), is selected, though for other responses types, loss functions such as the Gini Index may be selected. For any particular node split under consideration using covariate $w$,  that partitions individuals into two groups ($\{N_{w, 0}, N_{w, 1}\}$) of sizes $n_{w,0}$ and $n_{w,1}$, we minimize the criterion $SSE(w)=
\sum_{c=0,1}\sum_{q = 1}^{n_{w, c}} (\hat{S}_q^\tau(t) - \overline{{S}_c})^2$,
where $\overline{{S}_c} = \frac{1}{n_{w, c}}\sum_{q=1}^{n_{w, c}}\hat{S}_{q,c}^\tau(t)$, where $c= 0, 1$.

  
 \item[2c.] (Stopping criteria and prediction from the grown tree) Repeat Step 2a and 2b until one of the following hyperparameters has been met: minimum node size ($\min\{n_{w,c},c=0,1\}$) is set to $40$ in this paper) is reached, or $SSE(w)$ does not decrease any further with more branches. This results in a single binary decision tree. For each terminal node (``leaf'') of the tree $C$, predict $P\{T_i(t)\geq \tau\}$ by the average of the outcome for observations in leaf $C$. Let the prediction be $\Tilde{P}_b(\bZ_{i}(t); \hat{\btheta}_b)$, where $\Tilde{\btheta}_b$ is the collection of terminal node predictions in the $b$-th tree, and $\bZ_i(t)$ is the covariate profile for subject $i$. 
\end{itemize}

\item[Step 3] (Ensemble prediction): The final prediction for an individual $i'$ with covariates $\bZ_{i'}(t)$ is given by the average of all the outputs from $B$ binary decision trees above: $ \Tilde{P}_{i'}\{T_{i'}(t) \geq \tau|Z_{i'}(t)\}= \frac{1}{B}\sum_{b=1}^B \Tilde{P}_b(\bZ_{i'}(t); \hat{\btheta}_b)$.

\end{itemize}


\begin{remark} The two-stage bootstrapping scheme has a few statistical and computational advantages consistent with results from \citet{hrfRef}: (1) subjects with many measurements are as likely to be included in the bootstrap sample as subjects with few measurements, making it unlikely that observations from subjects with longer follow-up time overpower subjects with shorter follow-up times in the tree-growing stage and (2) out-of-bag data for each tree remains independent of the bagged data, making predictions from such a forest remarkably simple.
\end{remark}






Permutation tests are applied to the out-of-bag sample to generate $p$-values assessing statistical significance for predictors used within the random forest algorithm while controlling for the remaining predictors. Each variable, $W$, is selected and the values of $W$ in the out-of-bag sample are permuted $d$ times, to obtain $\mathscr{W}_s$, $s = 1, 2, ... d$. This breaks their association with $\hat{S}_i^\tau(t)$. The out-of-bag samples corresponding to both $\mathscr{W}_s$ and $\mathit{W}$ are then passed through the forest to obtain new predictions, $\mathcal{P}_{\mathscr{W}, s}$ and $\mathcal{P}_\mathit{W}$ respectively. Exploiting the fact that ${p}_{w,\text{test}}= \dfrac{\frac{1}{d}\sum_{s = 1}^{d} (\mathcal{P}_W-\mathcal{P}_{\mathscr{W}, s})}{\sigma _{\mathscr{W}, \mathit{W}}}$ has an asymptotically standard normal distribution, one can then obtain a Wald-type test of variable significance ($\sigma _{\mathscr{W}, \mathit{W}}$ is estimated from the permutation distribution of prediction errors).

\vspace{-1cm}
\section{Evaluation Metrics}
\label{s:evaluation}
Both the m-XMT models and \texttt{RFRE.PO} may be evaluated  using Harrell's C-statistic (also called C-index or the index of concordance), a non-parametric measure of concordance \citep{HarrellsCStat}. Briefly, the C-statistic compares the predicted risk scores or fit values for any two observations with the actual time-to-event for those two observations. The proportion of observations that are correctly ordered is reported as the C-statistic. Event pairs where censoring obscures the correct ordering of events within the pair are not included in the C-statistic calculation. Any model that perfectly orders patients based on some risk score has a C-statistic of 1, while a model with no predictive power has C-statistic equal to 0.50. C-statistics for events in the censored longitudinal data structure given in Section \ref{s:model} can be reported across all events or within $\tau$-duration follow-up windows starting at $t$. 
In addition to C-statistics, overall and time-dependent mean squared error (MSE) estimates are used to assess model calibration as described in Supplemental Materials Section \ref{s:supplementMSE}.
\vspace{-1cm}
\section{Simulation Study} 
\label{s:SimStudy}
As mentioned in Section \ref{s:intro}, random forest algorithms are known for nonparametrically capturing complex relationships between a large number of predictors and the outcome of interest, where these complex relationships might involve higher-order interactions, nonlinearity, and collinearity between predictors. In this section, we evaluate the ability of \texttt{RFRE.PO} 
to predict $P\{T_i(t)\geq \tau \mid \bZ_i(t)\}$ in a scenario where the true relationship between $\bZ_i(t)$ and $I\{T_i(t) \geq \tau\}$ is many times more complicated than a modeler who tends to use main effect terms would anticipate. For comparison, we also describe results from using a perfectly specified modified XMT model, labeled m-XMT (True), and a modified XMT model that only uses main effect terms for  $\bZ_i(t)$ in equation (\ref{mod-XMT}), labeled m-XMT (Main Effects).  

The goals of our simulation study are (1) to verify high performance of our random forest algorithm for prediction of $P\{T_i(t) \geq \tau|\bZ_i(t)\}$ in the recurrent event setting and (2) to demonstrate the importance of event rate history in modeling $P\{T_i(t) \geq \tau|\bZ_i(t)\}$.
As part of goal (2), for each modeling strategy [\texttt{RFRE.PO}, m-XMT (True), m-XMT (Main Effects)], we consider three different settings corresponding to availability of historical event rate information as described in Section \ref{s:model}: (1) \textit{none}, (2) $H_p^{(1)}(t)$, or (3) $H_f^{(1)}(t)$.

We now describe the details of how data were simulated. The true relationship between recurrent events ($T_{i, 1}< T_{i, 2}<\ldots$) and patient characteristics (continuous covariates $Z_1$, $Z_2$, $Z_4$, and categorical covariates $Z_3$, $Z_5$, $Z_6$, $Z_7$) is designed to be intentionally complex. 
Recurrent event gap-times marginally follow an exponential distribution with hazard: $\lambda_i =  \exp\{Z_{i2} \sin(Z_{i1}/Z_{i6}) + (-1)^{I(Z_{i2}> 2)}Z_{i3} +Z_{i1} Z_{i6} + Z_{i2}^2Z_{i4}\}.$ We simulate $Z_1, Z_2 \sim \text{MVN} (\mathbf{\mu}, \Sigma)$, where $\mathbf{\mu} = (0, 2)^\top$ and $\Sigma = \begin{pmatrix}
    5 & 0.7\\
    0.7 & .8
\end{pmatrix}$, $ Z_3 + 2 \sim \text{Poisson} (4)$, $Z_4 -0.1 \sim \text{Beta}(7, 1) $, $  Z_5  \in \{0, 1, 2\}$ with probability $ \big (\frac{2}{3}, \frac{1}{6}, \frac{1}{6}\big )$, $ Z_6  \in \{-5, -2, 2, 3\}$ with probability $ \big (\frac{1}{10}, \frac{1}{3}, \frac{11}{30}, \frac{1}{5}\big ),$ and $ Z_7  \in \{0, 1, 2, 3\} $ with equal probability. The sign in front of $Z_3$ depends on the value of $Z_2$ which is correlated with $Z_1$. $Z_5$ and $Z_7$ are noise covariates. Individuals with simulated hazards corresponding to mean times-to-recurrent event outside of the range of $\frac{1}{15}$ and $\frac{15}{8}$ years are removed from the sample and replaced. We induce correlation among exponentially distributed gap times for each subject $i$ by utilizing the Gaussian copula method as described by \citet{Xia:StatMed:2019:regression}, for $\rho = 0, 0.3, 0.6$ or $ 0.9$. The resultant recurrent event times are converted to a censored longitudinal data format as described in Section \ref{s:model}. Censoring is independently generated from an exponential distribution so that 0\%, $23\%$, $45\%$, or $63\%$ of individuals are censored.

The censored longitudinal data structure assumes $\mathcal{T} = \{0, \frac{1}{12}, \frac{1}{6},...2\}$\text{ years}, and $\tau = \frac{1}{6}$ year. For individual $i$, $\mathcal{T}_i $ is the subset of follow-up window start-times in $\mathcal{T}$ where individual $i$ remains at risk for recurrent events. Both the m-XMT (True) model and the m-XMT (Main Effects) estimate $\text{logit} \{E(S^\tau _i(t))\}$ assuming the relationship $\text{logit} \{E(S^\tau _i(t))\} = \mathbf{\beta} ^\top \bZ^*_i(t)$. For m-XMT (True), the ad hoc nonlinear model, $\bZ^*_i(t)= \{ Z_{i 2} \sin(Z_{i 1}/Z_{i 6}),$ $ (-1)^{I(Z_{i2}> 2)}Z_{i3},$ $
Z_{i1}Z_{i6},$ $
 Z_{i2}^2Z_{i4}\}^\top$. The m-XMT (Main Effects) model has covariate vector $\bZ^*_i(t) = \{Z_{i1}, $ $Z_{i2}, $ $Z_{i 3}, $ $Z_{i4},$ $Z_{i5},$ $ Z_{i6},$ $Z_{i7}, t\}^\top$. Covariate histories are included as additional patient characteristics in $\bZ^*_i(t)$ for full and partial history availability settings. For full history availability settings this predictor takes the form $H_{f,i}^{(1)}(t)$ as given in Equation (\ref{eq:fullZ}) with $\mathcal{T}^*_i(t)=\{\frac{-1}{12}\} \cup \{t_k \in \mathcal{T}_i : t_k+\tau^* < t\}$, and $\tau^* = \frac{1}{12}$. For  partial history availability settings, imputes for missing event histories prior to $t=0$ follow recommendations in Section \ref{s:model} with $\Tilde{T}_{i,j} \sim $Exp$(\lambda_i)$, where $\lambda_i \sim$Unif$(\frac{8}{15}, 15)$. Partial history covariates are added to $\bZ^*_i(t)$ based on Equation (\ref{eq:partialZ}), with the same $\mathcal{T}^*_i(t)$ and $\tau^*$ as $H_f^{(1)}(t)$. All finite sample performance results are based on 500 simulated datasets of n=500 participants per scenario.

\begin{figure}
    \centering \includegraphics[height=.8\textheight]{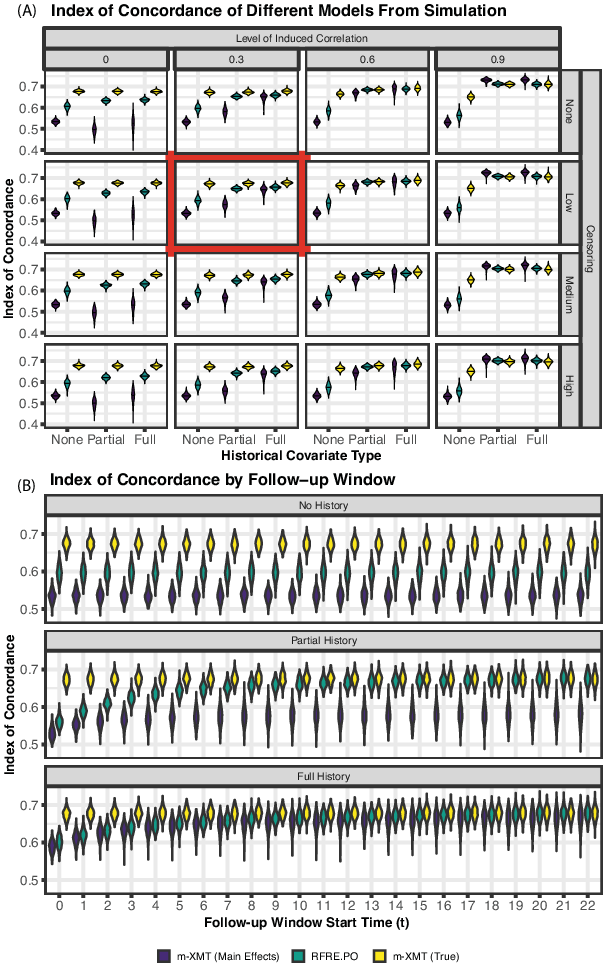}
    \caption{Simulated model fit for (1) the modified XMT model using main effects terms only \{m-XMT (Main Effects), shown in purple\}, (2) our random forest algorithm for recurrent events (RFRE.PO, shown in green) and (3) the true modified XMT model \{m-XMT (True), shown in yellow\}. Results are based on 500 simulated datasets of n=500 participants per scenario. Panel (A) displays violin plots of C-statistics (indices of concordance) seen in simulation. For partial history information, average C-statistics from $m=10$ imputed data sets are shown. Measures of center and spread are in Table \ref{tab:simResultTable}. Within a box, historical covariate information increases from left to right.  Panel (B) expands on the scenario boxed in red from panel (A), i.e., the low censoring setting with 0.3 correlation. Here violin plot C-statistic distributions are time-dependent,  reflecting model fit for $\tau=\frac{1}{6}$ duration follow-up windows starting at times, $t$, marked on the horizontal axis. The quality of historical information increases from top to bottom. 
    }
    \label{fig:SimPlot}
\end{figure}

Figures \ref{fig:SimPlot}A and  \ref{fig:SimPlot}B display simulated model fit for (1) the modified XMT model using main effects terms only \{m-XMT (Main Effects), shown in purple\}, (2) our random forest algorithm for recurrent events (RFRE.PO, shown in green) and (3) the true modified XMT model \{m-XMT (True), shown in yellow\}. Panel (A) displays violin plots of C-statistics (indices of concordance) seen in simulation with corresponding measures of center and spread shown in Table \ref{tab:simResultTable}. For partial history information, average C-statistics from $m=10$ imputed data sets are shown.  Within a box, historical covariate information increases from left to right.  Panel (B) expands on the scenario boxed in red from panel (A), i.e., the low censoring setting with 0.3 correlation. Here violin plot C-statistic distributions are time-dependent,  reflecting model fit for $\tau=\frac{1}{6}$ duration follow-up windows starting at times, $t$, marked on the horizontal axis. The quality of historical information increases from top to bottom. 
 The m-XMT (True) model performance provides a benchmark for assessing performance of the proposed (truth-blind) dynamic prediction methods, the m-XMT (Main Effects) model and the \texttt{RFRE.PO} algorithm.

Panel A results show that when there is no historical covariate used for dynamic prediction, the m-XMT (True) model has the best C-statistic performance, as expected, for all levels of censoring and correlation, followed by the \texttt{RFRE.PO} algorithm and then the  m-XMT (Main Effects) model.   As correlation between recurrent event times increases, use of either partial or full history covariates in these latter two dynamic prediction algorithms brings C-statistic performance closer to that of the m-XMT (True) model. When using full history covariates,  \texttt{RFRE.PO} outperforms the m-XMT (Main Effects) model for $\rho=0.3$ and performs similarly for  $\rho=0.6$. When using partial history covariates, \texttt{RFRE.PO} outperforms m-XMT (Main Effects) for both $\rho=0.3$ and $\rho=0.6$. For  $\rho=0.9$, use of either partial or full history covariates in the main effects model slightly outperforms both the \texttt{RFRE.PO} algorithm and the m-XMT (True) model.  

Generally, the inclusion of full history covariates is more effective than that of partial history covariates, though the benefit of having full history covariates available is most strongly seen in the lowest two correlation settings when applying the m-XMT (Main Effects) model.  In the  highest correlation setting, partial and full covariate history algorithms have very similar performance. In the absence of correlation between recurrent event times, use of history covariates shows modest, if any, improvement for the \texttt{RFRE.PO} algorithm C-statistic whereas C-statistic performance for the m-XMT (Main Effects) model deteriorates noticeably. 


Time-dependent C-statistic performance highlighted in Panel B shows how quickly algorithms using partial and full covariate history information improve and stabilize over the course of follow-up. Stabilization happens quickly, with little change in time-dependent follow-up windows after $t=5$. 
Because covariate history variables improve with increasing $t$, there is an induced interaction between these predictors and $t$, particularly when the algorithms are using the partial history covariates. This type of interaction is naturally assessed by the \texttt{RFRE.PO} approach.   

Together, Figures \ref{fig:SimPlot}A and  \ref{fig:SimPlot}B  indicate that in highly correlated settings, (1) patient history is an essential predictor to include in dynamic prediction algorithms and (2) the \texttt{RFRE.PO} algorithm seems to outperform the m-XMT (Main Effects) algorithm in all but the very highest correlation setting where history covariates are used for dynamic prediction. Additional simulation results are available in Supplemental Materials. Section \ref{s:supplementMSEInterp} summarizes mean squared error results with corresponding Table \ref{tab:simMSETable}, while Section \ref{s:supplementImportance}  describes the ability of these algorithms to distinguish signal covariates ($Z_1,Z_2,Z_3,Z_4,Z_6$) versus noise covariates ($Z_5,Z_7$) in Tables \ref{tab:m-XMTSig} and \ref{tab:checkPower}.

\vspace{-1cm}
\section{Prediction of COPD Exacerbations}
\label{s:data-app}
Chronic obstructive pulmonary disease (COPD) is most often attributed to smoking history, although cases may also be caused by breathing polluted air such as biomass fuel or industrial chemicals \citep{copdBurden}. COPD is characterized by periods of relative stability punctuated with episodic exacerbations of coughing, struggling for breath, and/or wheezing that may lead to hospitalization, in which case exacerbations are called \say{severe.} Such exacerbations increase overall disease progression, eventually leading to lung transplantation or death
\citep{ExacerbationEffect, LungFunction}, indicating some correlation between events within a subject. It is of clinical interest to dynamically predict the chances of being exacerbation-free over time.

Azithromycin for the Prevention of COPD \citep{azithro} was a multi-center clinical trial evaluating the effect of daily oral azithromycin (250mg) in preventing severe COPD exacerbations during a year of follow-up. We apply \texttt{RFRE.PO} to predict the probability of remaining exacerbation-free over a subsequent $\tau= 180$ day follow-up window in this cohort ($n=1035$ study participants with complete baseline data and longitudinal sleep study variables).
There are 962 available covariates, including demographic variables, social well-being surveys, sleep quality metrics and clinical characteristics such as forced vital capacity (FVC), and medication use. There is strong correlation between variables as some are derived from each other, making \texttt{RFRE.PO} an attractive alternative to model-based predictions (see Supplemental Figure \ref{fig:corrHeatMap}). Censored longitudinal data is constructed with $\mathcal{T}= \{0, 30, 60,..., 180\}$, and used for all analyses, with $T_i = \{0, 30, \ldots t_{b_i}\}$ for individual $i$.

Baseline information on recurrent event times prior to $t=0$ were not available. We defined a full history covariate, $H_{f,i}(t)$, as $I$(Severe exacerbation recorded prior to $t$), with $H_{f,i}(0)$=0. We also considered two partial covariate history variables from Section \ref{s:model}, the categorical version of $H_{p,i}^{(2)}(t)$ given in Equation \ref{eq:fullTimeSince}, with $H_{p,i}^\dag(t)$ replacing $H_{f,i}^\dag(t)$, and $H_{p,i}^{(3)}(t)$.  For categorical $H_{p,i}^{(2)}(t)$, we assume $\tau^*= 1$ year and $\mathcal{T}^*_i(t) = \{t^* \in \mathcal{T}_i \cup -1 \mbox{ year}  : t^* + 1 \mbox{ year} \leq t\}$. This variable categorizes $H_{p,i}^\dag(t)$ in the same manner Equation \ref{eq:fullTimeSince} categorizes $H_{f,i}^\dag(t)$ with higher values indicating shorter exacerbation-free periods leading up to $t$. For $H_{p,i}^{(3)}(t)$, we assume $\tau^*=30$ days and $\mathcal{T}^*_i(t) = \{\mathcal{T}_i \cup -30 \mbox{ days}: t^* + 30 \mbox{ days }\leq t\}$, so that this variable estimates the  probability that prior follow-up windows had an exacerbation within 30 days. Imputed event times prior to time $0$ were sampled from an exponential($\lambda_i$) distribution with $1/\lambda_i$ generated from a uniform(100 day, 180 day) distribution, which was comparable to placebo group outcomes. Rubin's rule \citep{RubinsRulesCite} was used to combine inference across multiply imputed datasets as described in Section \ref{s:model}.

For comparison purposes, we present \texttt{RFRE.PO} results alongside results from two m-XMT models.
The first m-XMT model (Clinical Input) includes baseline variables used by \citet{Xia:StatMed:2019:regression} for this cohort; these predictors are commonly used to assess confounder-adjusted treatment effects in COPD journals.  The second m-XMT (Forward Selection) model is based on Wald forward selection applied to the first multiply imputed data set with $p< 0.05$ required for entry; covariates that introduce model instability are discarded from consideration. 
Prediction algorithms for \texttt{RFRE.PO} and the forward selection model were applied to a training cohort (70\% of original cohort) and all algorithms were evaluated via the C-statistic in a validation cohort (remaining 30\% of original cohort). For any particular multiply imputed validation dataset, standard errors of the C-statistic are calculated via the bootstrap with $b=100$ bootstrap samples. C-statistics for models involving multiply imputed history variables are combined across multiply imputed datasets using Rubin's rule. Parameter estimates and standard errors for m-XMT (Forward Selection), which involves the imputed partial history covariates, are combined across multiply imputed validation datasets via Rubin's Rule. Permutation tests associated with the \texttt{RFRE.PO} algorithm predictors are calculated within the training dataset, using out-of-bag samples as described in Section \ref{s:HRF}.

For m-XMT (Clinical Input), where estimation of treatment effect, and not necessarily prediction was the goal, predictors 
are baseline forced expiratory volume in one second in liters (Baseline FEV$_1$), age in decades (Age10), male gender \{$I$(Male)\}, and baseline smoking status \{$I$(Smoker)\}:
\begin{equation*}
    \text{logit}\{E({S}^\tau _{i}(t)) \}= \beta_0 +  \beta_1 I(\text{Treatment})_i +\beta_2 (\text{Baseline FEV$_1$})_{i}+\beta_3 (\text{Age10})_i + \beta_4 I(\text{Male})_i+\beta_5 I(\text{Smoker})_i,
\end{equation*} where parameter estimates for the m-XMT model are displayed in Supplemental Table \ref{tab:summer-estimates}. 


The m-XMT model based on Wald forward selection, is: 
\begin{align*}
      & \text{logit}\{E({S}^\tau _{i}(t) )\}  = \beta_0 + \beta_1 I\{\text{Categorical } H_{p, i}^{(2)}(t) = 0\} +  \beta_2I(\text{Hospitalization in year before $t=0$})_i \\
      & \quad +
      \beta_3 I(\text{Corticosteroid use in year before $t=0$})_i+
      \beta_4 H_{f,i}(t)+
      \beta_5 t +
      \beta_6 I(\text{Male})_{i}+
      \beta _7 I(\text{Race = Black})_i\\
      & \quad  + 
     \beta_8 I(\text{Sudden feelings of panic at time } t)_{i}+
     \beta_9 I(\text{Sudden feelings of \say{butterfly} fright at time }t)_i\\
     &\quad +
     \beta_{10} I(\text{Seen a mental health provider at time }t)_{i} + 
     \beta_{11} I(\text{Has pain at time }t)_i\\
     &\quad  +
     \beta_{12}(\text{St. George Respiratory Questionnaire symptom score at time } t)_{i}+
     \beta_{13}I(\text{Antiarrhythmic use at time }t)_{i} \\ 
     &\quad+ 
     \beta_{14} (\text{FEV}_1 \text{/FVC \% predicted at time } t)_{i}+ 
     \beta_{15} I(\text{Azithromycin treatment group})_{i} \\
     &\quad + 
     \beta_{16} I(\text{Inhaled corticosteriod or long-acting beta-agonist use at time }t)_{i}+ 
     \beta_{17}I(\text{Other baseline medications})_i \\
     &\quad+ 
     \beta_{18} (\text{Supplemental oxygen use at time } t)_{i}+ 
     \beta_{19} I (\text{Mixed beta-blocker use at time }t)_i\\
     &\quad +
     \beta_{20} I(\text{Bronchiectasis diagnosis at time }t)_{i}
 \end{align*}
 with parameter estimates displayed in Supplemental Table \ref{table:wald-estimates}. 

For the \texttt{RFRE.PO} algorithm, Supplemental Table \ref{table:perm-tests} highlights predictors that had a statistically significant permutation test Z-score for at least one of the multiply imputed training datasets. Additional predictors that did not achieve statistical significance via the permutation test, but were used in the \texttt{RFRE.PO} algorithm in some form, are not shown.

\begin{figure}
    \centering \includegraphics[width=\textwidth]{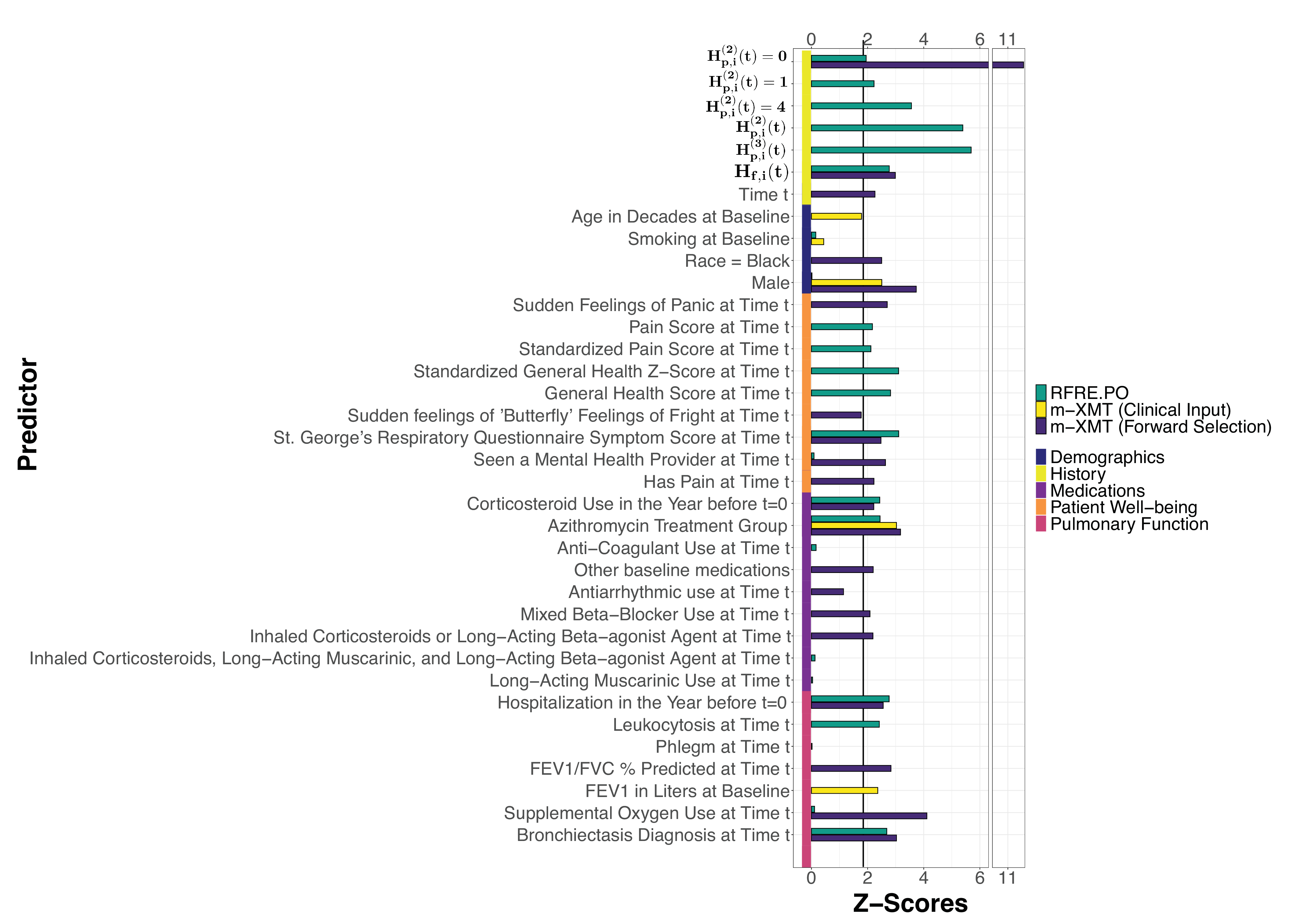}
    \caption{Wald Statistic Z-scores derived from the validation set for predictors that were included in the m-XMT models ( Clinical Input model  from Xia, Murray, Tayob 2020) and the m-XMT  selected based on forward selection of the Wald statistic, after adjusting for multiple imputation using Rubin's Rule, where appropriate, and permutation test Z-scores derived from out-of-bag samples as part of the \texttt{RFRE.PO} algorithm, adjusting for multiple imputation using Rubin's Rule where appropriate. For the \texttt{RFRE.PO} method, 100 permutations were used in calculating the permutation Z-scores. Note that some Wald forward selection variables lost significance as additional terms were added. Colors along the y-axis characterize the type of predictor displayed. Predictors are grouped into one of five categories, displayed in the legend on the right. For the analysis, we created three history variables, $H_f(t), H_{p}^{(2)}(t)$ and $H_{p}^{(3)}(t)$. $H_f(t)$ was an indicator if the subject had experienced a severe exacerbation on study time prior to time $t$. $H_{p}^{(2)}(t)$ was assumed to take a categorical form based on whether the participant had an exacerbation in the past zero to 31 days ($H_{p}^{(2)}(t) =4$), 31-92 days ($H_{p}^{(2)}(t) =3$), 93-182 days ($H_{p}^{(2)}(t) =2$), 183 days to 365 days ($H_{p}^{(2)}(t) =1$), or finally, if that time was more than 365 days or never ($H_{p}^{(2)}(t) =0$). $H_{p}^{(3)}(t)$ is the estimated 30-day event rate using all follow-up prior to time $t$. Both of these partial history variables are updated at later follow-up windows starting at $t>0$ based on observed exacerbation data for each individual.}
    \label{fig:zBars}
\end{figure}

Figure \ref{fig:zBars} displays validation dataset Z-scores for predictors included in the m-XMT (Clinical Input) and  m-XMT (Forward Selection)  algorithms as well as out-of-bag sample Z-scores 
for \texttt{RFRE.PO} predictors highlighted in Supplemental Table \ref{table:perm-tests}. In algorithms that included history covariates \{m-XMT (Forward Selection) and \texttt{RFRE.PO}\}, these predictors showed strong validation results in Figure \ref{fig:zBars}.  When prediction algorithms were applied to individuals in the validation cohort, the \texttt{RFRE.PO} algorithm outperformed both the m-XMT (Clinical Input) and m-XMT (Forward Selection) algorithms in terms of both the C-statistic (C-statistic $0.613$ 
compared to $0.564 $ 
and $0.571$, 
respectively) and estimated mean squared-error (MSE $0.242 $ 
compared to $0.252 $ 
and $0.264) $; See Table \ref{table:data-res}.  
Time-dependent C-statistics and $MSE(t)$ for follow-up windows starting at time, $t$, in the validation dataset are displayed in Figure \ref{fig:TimeVaryingCMSE} for the various prediction algorithms. Compared to the other algorithms, \texttt{RFRE.PO} has fairly stable and attractive performance metrics across the various follow-up windows, likely because interactions with the various predictors over time are naturally incorporated if useful for prediction.  

\begin{figure}
    \centering \includegraphics[height = .7 \textheight]{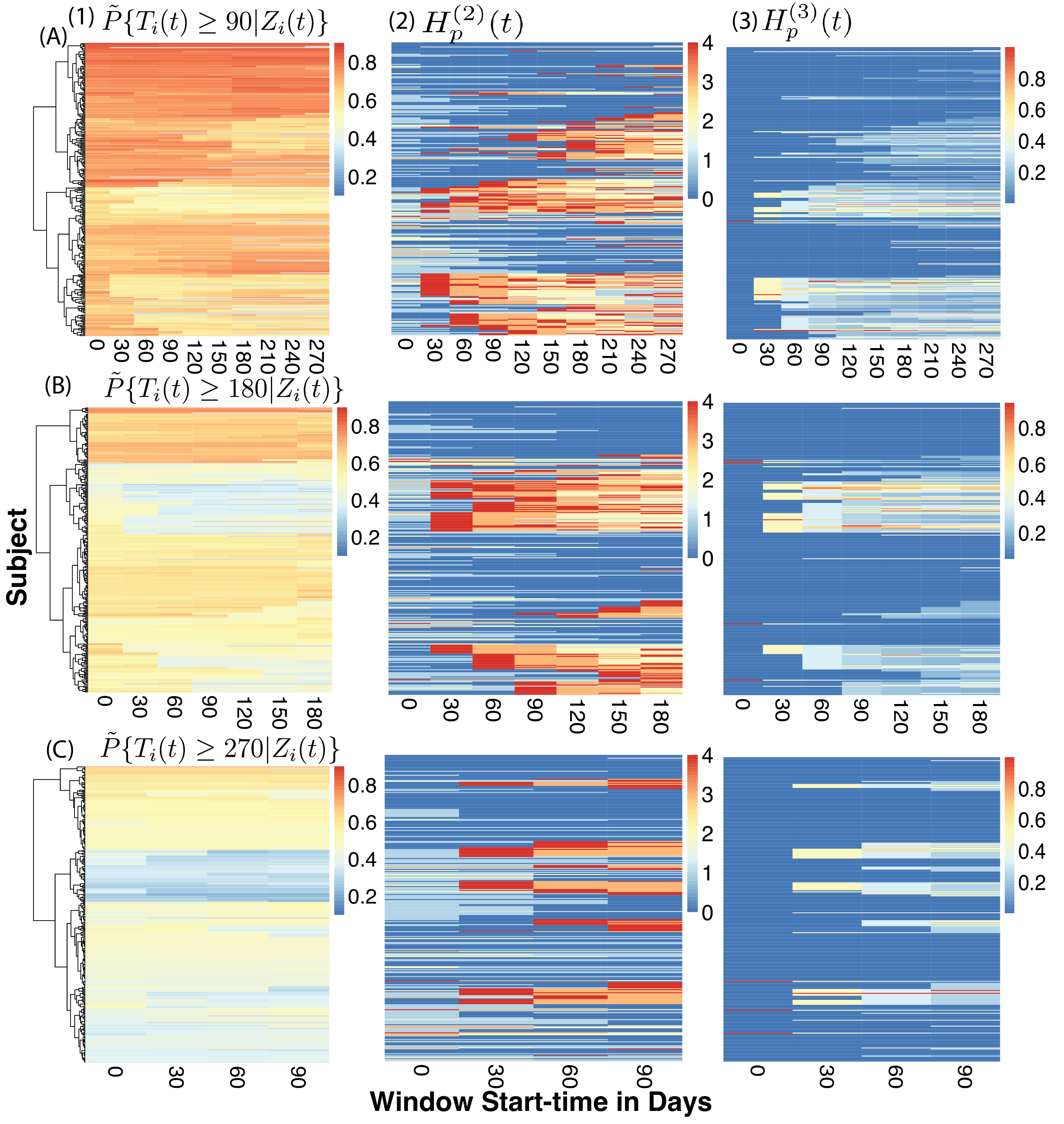}
    \caption{\texttt{RFRE.PO} predictions and history covariates from the Azithromycin for the Prevention of COPD validation cohort via a $3\times 3$ panel heatmap display. Within each row of heatmaps, subjects are ordered vertically, and clustered according to the default settings of the \texttt{pheatmap} command in R applied to the fitted dynamic prediction values for that row of heatmaps. Heatmap rows marked (A), (B) and (C) vary values of follow-up window duration, $\tau$, assuming $\tau=90$, $\tau=180$ and $\tau = 270$, respectively. Heatmap columns marked (1), (2) and (3) display: (1)  dynamic predictions, $\tilde{P}\{T_i(t) \geq \tau | Z_i(t)\}$, shown for follow up windows starting at $t$, with $t$ spaced every 30 days until $t+\tau$ exceeds 360, (2) corresponding categorical $H_{p}^{(2)}(t)$ and (3) corresponding continuous $H_p^{(3)}(t)$. Both partial history variables are described in detail in Section \ref{s:model}, and the specification of $\mathcal{T}^*_i(t)$ and $\tau^*$ are described in Section \ref{s:data-app}.
    Although individuals are not clustered identically in the heatmaps shown in rows (A), (B) and (C), the heatmap color schemes (as expected) reflect higher probabilities of remaining event-free over shorter follow up windows, i.e., $\tilde{P}\{T_i(t) \geq 90 | Z_i(t)\}$ $>$ $\tilde{P}\{T_i(t) \geq 180 | Z_i(t)\}$ $>$ $\tilde{P}\{T_i(t) \geq 270 | Z_i(t)\}$.
    The display of the history variables seems to reflect in the fitted values from the \texttt{RFRE.PO} method, as patterns in column (1) are roughly represented in columns (2) and (3).}
    \label{fig:appPanelGraphic}
\end{figure}

As can be seen in Figure \ref{fig:zBars}, variables identified via m-XMT models are for the most part identified as important variables in the \texttt{RFRE.PO} algorithm, although the version of the predictors may vary. For example, variables related to the long-acting muscarinic or beta-blockers, and/or steroid treatment are incorporated in different forms into these models (highlighted on the y-axis in purple). Similarly, different variables relating to patient well-being (highlighted in orange) are featured in the m-XMT (Forward Selection) model (has pain at time $t$) versus the \texttt{RFRE.PO} algorithm (pain score at time $t$ and standardized pain score at time $t$), as well as different measures of pulmonary function (highlighted in pink). The \texttt{RFRE.PO} algorithm highlights a few variables that were not used in the forward selection model, including general health score at time $t$ and leukocytosis present at time $t$. The \texttt{RFRE.PO} algorithm has no problem including information from correlated predictors; for instance \texttt{RFRE.PO} included all history variables, whereas this introduced model instability in the m-XMT (Forward Selection) model. Figure \ref{fig:appPanelGraphic} displays a $3 \times 3$ panel heat map of survival probabilities with corresponding history covariates. Within a row of panels, the general scheme of the predicted values are reflected in the values of derived history covariates, indicating the utility of these covariates in creating dynamic predictions for the Azithromycin for the Prevention of COPD cohort. 


The \texttt{RFRE.PO}  algorithm includes predictors, with corresponding split decisions, that minimize squared error of the predictions relative to the pseudo-observations. 
This algorithm is completely agnostic to statistical significance and scientific experience, so long as the squared error within nodes along cut-points decreases, so validation checks are quite important. 
With variable power to detect statistical significance of useful predictors,  methods that require statistical significance for model inclusion may be disadvantaged in terms of prediction when compared to the \texttt{RFRE.PO} algorithm that does not have this requirement.  
And although the \texttt{RFRE.PO} algorithm  may occasionally include a predictor that is not useful in the validation set (type I error analog), it has a much better chance of including predictors that are useful but would not have met statistical significance (akin to a type II error) in traditional statistical model building environments. 
If the scientific goal of analysis is prediction, the \texttt{RFRE.PO} approach seems remarkably simple and effective.
\vspace{-1cm}
\section{Discussion}
\label{s:discussion}
We presented two novel approaches, \texttt{RFRE.PO} and the m-XMT model, for dynamically estimating probabilities of being event-free across a sequence of follow-up windows.  An additional novelty is our advice for incorporating covariate history into  prediction algorithms, with a multiple imputation approach when covariate history is not known prior to time $t=0$. For purposes of prediction, \texttt{RFRE.PO} tends to outperform the m-XMT (Main Effects) model that a traditional modeler might first attempt. However, the m-XMT model has the advantage of interpretable parameters and a clear formula for prediction that offers more transparency to users, and may accommodate further interactions than explored in the m-XMT (Main Effects) model.  


Our presentation  did not touch on how causal relationships might be evaluated using these methods.  In the Azithromycin for Prevention of COPD study, assessment of the treatment effect disallows using information on time-dependent predictors that could have been changed by treatment after time $t=0$.  To do so would adjust away part of the treatment effect under study. Similarly, clinical trial analyses may require adjustment for baseline confounders even if they do not improve prediction metrics. The m-XMT model allows for more intentional use of covariates to match scientific goals of analysis when compared to the \texttt{RFRE.PO} algorithm.  The m-XMT (Clinical Input) model was the only appropriate model for assessing the effect of azithromycin presented in this manuscript.

The growing popularity of machine learning tools is likely to dominate analyses of high-dimensional data, and the \texttt{RFRE.PO} algorithm contributes to this literature as one of the first machine learning methods to be able to make dynamic predictions from recurrent event data. There are tuning parameters in random forest algorithms that may potentially be tuned for better performance, warranting further research.  
Some authors have studied parameter tuning \citep[e.g.,][]{tuneRF3}, but we have not seen work for longitudinal random forests to date. The development of (1) the theory of tuning parameters for longitudinal random forests and (2) the potential application of the censored longitudinal data structure for dynamic prediction to other machine learning algorithms are all important future research areas.

\section*{Supplementary Material}
Supplementary material is available online at \url{http://biostatistics.oxfordjournals.org}.

\vspace{-.5cm}
\section*{Acknowledgment}                           

\vspace{-.25cm}                                                          
The authors would like to thank the COPD Clinical Research Network, particularly our University of Michigan collaborator, Meilan Han, for the use of their data. Additional appreciation is given to the Editors, Associate Editor, and two anonymous reviewers whose comments significantly improved the presentation of the revised paper. The research is supported in part by a Precision Health Investigator Award from University of Michigan, Ann Arbor (AL, ZW). \textit{Conflict of Interest}: None.

\vspace{-.5cm}
\section*{Data Availability Statement}          

\vspace{-.25cm}                  
Code for simulation and data analysis is available at \url{https://github.com/AbigailLoe/RFRE.PO}. The data that support the findings in this paper are available from the corresponding author upon reasonable request.

\vspace{-.5cm}
\bibliographystyle{biorefs}
\bibliography{rfrepo}
\label{LastPage}

\newpage

\end{document}


\def\spacingset#1{\renewcommand{\baselinestretch}%
		{#1}\small\normalsize} \spacingset{1}
	
\if0\blind
	{		
\title{\bf Supplemental Materials for \\``Random Forest for Dynamic Risk Prediction of Recurrent Events: A Pseudo-Observation Approach"}
		\author[1,$^\ast$]{Abigail Loe}
		\author[1,$^+$]{Susan Murray}
		\author[1,$^+$]{Zhenke Wu}
		\affil[1]{Department of Biostatistics, University of Michigan, Ann Arbor, MI 48109, USA; E-mail: $^\ast${\tt agloe@umich.edu}; $^+$: co-senior authors}
		
		\date{}
		\maketitle
} \fi
	
\if1\blind
	{
		\title{\bf Random Forest for Dynamic Risk Prediction of Recurrent Events: A Pseudo-Observation Approach}
		\author[]{}
		\date{}
		\maketitle
	} \fi
	
\vspace{-1cm}

\bigskip




\renewcommand{\thefigure}{S\arabic{figure}}
\renewcommand{\theequation}{S\arabic{equation}}
\renewcommand{\thetable}{S\arabic{table}}
\renewcommand{\thetheorem}{S\arabic{theorem}}
\renewcommand{\thesection}{A\arabic{section}}

\setcounter{figure}{0}
\setcounter{equation}{0}
\setcounter{table}{0}
\setcounter{theorem}{0}

\section{Kaplan-Meier Estimation}\label{s:KMprimer}
Using the notation from Section \ref{s:model}, let $t \in \mathcal{T}$, and define:
\begin{align*}
    N_t(\tau)&= \sum_{i=1}^{n(t)} I[X_i(t)\leq \tau, \delta_i(t) =1]\\
    dN_t(u) &= \sum_{i=1}^{n(t)}I[X_i(t) = u, \delta_i (t) = 1]\\
    Y_t(u) &= \sum_{i=1}^{n(t)}I[X_i(t) \geq u]
\end{align*}  where $n(t)$ is the number of subjects available for follow-up at time point $t$. Briefly, $N_t(\tau)$ counts the number of events before horizon $\tau$, $dN_t(u)$ counts the number of events at precisely $u$, and $Y_t(u)$ keeps track of the number of subjects at-risk for a recurrent event at time point $u$. Then the Kaplan-Meier estimator for probability of survival greater than $\tau$ at $t\in \mathcal{T}$ is defined as
$$
\hat{P}[T_i(t) \geq \tau] = \prod_{i: X_{i}(t) \leq \tau}\Big\{1- \dfrac{dN_t[X_i(t)]}{Y_t[X_i(t)]} \Big\}.
$$
This probability can be estimated for each $t \in \mathcal{T}$ using the quantities calculated in Section \ref{s:model}. 

\section{Model Calibration via Mean Squared Error}\label{s:supplementMSE}
A calibration estimate of the m-XMT predictions averaged over all individuals, $i=1,\ldots,N$ and follow-up windows within individual, $t_0,\ldots,t_{b_i}$ becomes $$MSE=\frac{1}{\sum_{i=1}^{N} (b_i+1) }\sum_{i=1}^N \sum_{j=0}^{{b_i}} [\hat{S}_i^{\tau}(t_j)-\hat{P}\{T_i(t_j)\ge \tau | \bZ_i(t_j)\}]^2.$$ To obtain a similar calibration estimate for \texttt{RFRE.PO}, we replace the fitted values from the m-XMT model, $\hat{P}\{T_i(t_j)\ge \tau | \bZ_(t_j)\}$, with the predictions from the \texttt{RFRE.PO} method, $\Tilde{P}\{T_i(t_j)\ge \tau | \bZ_i(t_j)\}$

A time-dependent MSE estimate pertaining to m-XMT predictions for the follow-up window starting at $t \in \mathcal{T}$ is:
$$ 
MSE(t)=\frac{1}{n(t)}\sum_{i = 1}^{n(t)} [\hat{S}_i^{\tau}(t)-\hat{P}\{T_i(t)\ge \tau | \bZ_i(t)\}]^2.
$$ Similarly, we can calculate $MSE(t)$ for \texttt{RFRE.PO} by replacing the predictions from an m-XMT model with the predictions from the \texttt{RFRE.PO} method, as described in Section \ref{s:HRF}.

\section{Additional Simulation Results: MSE } \label{s:supplementMSEInterp}
\begin{figure}
    \centering \includegraphics[height=.9\textheight]{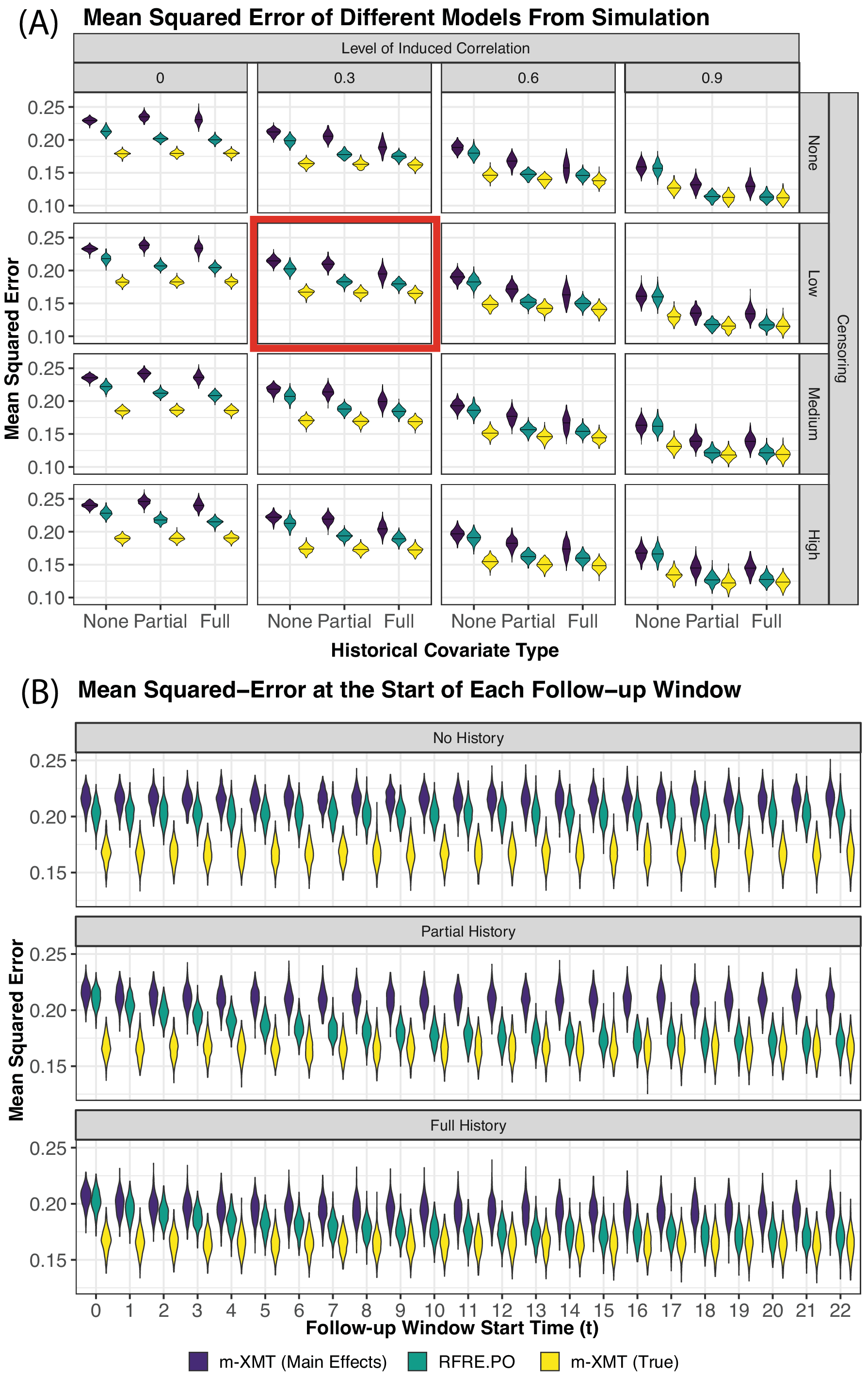}
    \caption{\footnotesize Panel (A) displays the mean squared error (MSE) for simulation settings displayed in Figure \ref{fig:SimPlot}. Censoring increases from top to bottom, and the amount of induced correlation increases from left to right. Within a box, the quality of historical information increases from left to right. Panel (B) shows the MSE at the start of each follow-up window for the setting boxed in red in Panel (A). From top to bottom, more historical information is added in Panel (B), and the MSE corresponding to \texttt{RFRE.PO} in $H_p^{(1)}(t)$ settings. 500 replicates were performed, each with $n=500$ subjects.}\label{fig:MSEplot}
\end{figure}

Figure \ref{fig:MSEplot} and Table \ref{tab:simMSETable}  summarize  $MSE$ and $MSE(t), t\in \mathcal{T}$, simulation results for the \texttt{RFRE.PO} algorithm, the m-XMT (Main Effects) model, and the m-XMT (True) model.  In general, the m-XMT (Main Effects) model has the highest MSE \{Panel (A)\}. In settings with correlation, the inclusion of history covariates decreases MSE for the m-XMT (Main Effects) model and the \texttt{RFRE.PO} algorithm. In all simulation settings, the \texttt{RFRE.PO} method outperforms the m-XMT (Main Effects) method, and upon the inclusion of history covariates, competes with the m-XMT (True) model in settings with highest correlation and lowest censoring. Examination of $MSE(t)$ reveals a similar stabilization as the stabilization observed in the C-statistic (Figure \ref{fig:MSEplot}).

\section{Additional Simulation Results: Power and Type I Error}\label{s:supplementImportance}

Tables \ref{tab:m-XMTSig} and \ref{tab:checkPower} display the simulated proportion of times predictors were deemed statistically important in dynamic prediction methods using the \texttt{RFRE.PO} algorithm (based on the random forest permutation test) and the m-XMT (Main Effects) model (based on the Wald test), respectively. Columns $Z_5$ and $Z_7$ of these tables show low rates of deeming noise covariates as important predictors with more conservative rates seen using the \texttt{RFRE.PO} algorithm than the m-XMT (Main Effects) model. For the m-XMT (Main Effects) model, these rates reflect type I error rates for inclusion of these predictors in the model that are reasonably close to the nominal 0.05 type I level subject to simulation variability. For the  \texttt{RFRE.PO} algorithm, these rates do not reflect the rate at which the predictors were used in the algorithm, but rather the rate at which these variables were significantly  improving prediction outputs when included as an input in the algorithm. 

Similarly, columns $Z_1, Z_2, Z_3, Z_4,$ and $ Z_6$
of these tables show the rate at which covariates associated with recurrent events in the m-XMT (True) model were deemed important in the \texttt{RFRE.PO} algorithm and the m-XMT (Main Effects) model in settings where (1) no history covariates were used, (2) $H_p^{(1)}(t)$ was included in the model or (3) $H_f^{(1)}(t)$ was used in the model.   For the m-XMT (Main Effects) model, these rates correspond to power for detecting the association of these predictors in these 3 covariate history settings. For the 
\texttt{RFRE.PO} algorithm, these rates reflect the rates at which these variables were significantly improving prediction outputs when included as an input in the algorithm in these three history covariate settings.

When no history covariates are included as algorithm inputs, the \texttt{RFRE.PO} algorithm is more likely to deem these predictors as significantly helpful in making dynamic predictions when compared to the m-XMT (Main Effects) model.  In settings where there is positive correlation between recurrent events and history covariates are included as inputs in the prediction algorithms, the relative importance of these history covariates tends to diminish rates of importance seen for $Z_1, Z_2, Z_3, Z_4, Z_6$ in Tables \ref{tab:m-XMTSig} and \ref{tab:checkPower}.


We were initially puzzled when looking at the importance of the covariate, $t$, in these tables, as $t$ was not directly used in generating data via the m-XMT (True) model. Yet, the covariate, $t$, tends to show high importance in models with covariate history variables and correlation between recurrent event times. The explanation seems to lie with Panel B of Figure \ref{fig:SimPlot}, where the C-statistic improves over the early follow-up windows when covariate history variables are used. This result seems to indicate an interaction between $t$ and history variables in settings with correlated recurrent events that were not explicitly modeled in the data generation stage.

\section{Simulation Study: Additional Tables and Figures}\label{s:supplementSimTabsFigs}

\begin{table}[ht]\footnotesize
 \caption{Simulated C-statistics listed as Mean (ESD) for varying prediction algorithms when $Z_1, Z_2$ are correlated with each other, censoring percentages (none is 0\%, light is 23\%, moderate is 45\%, and heavy is 63\%), and gap-time correlations ($\rho = 0$, $0.3$, $0.6$, $0.9$). 500 replicates were performed with $n=500$ for each replicate, and the same data is reported in Figure \ref{fig:SimPlot}.} \label{tab:simResultTable}
 {\begin{tabular*}{\textwidth}{@{}l@{\extracolsep{\fill}}c@{\extracolsep{\fill}}c@{\extracolsep{\fill}}c@{\extracolsep{\fill}}c@{\extracolsep{\fill}}c@{}}
 \hline 
 & $\rho = 0$ & $\rho = 0.3$ & $\rho = 0.6$ & $\rho = 0.9$ \\ 
 \hline 
 & & No Censoring \\  
 \hline 
 m-XMT (Main Effects) & 0.534 (0.009) & 0.533 (0.010) & 0.533 (0.010) & 0.531 (0.011)\\  
 \texttt{RFRE.PO} & 0.605 (0.012) & 0.596 (0.014) & 0.584 (0.015) & 0.563 (0.016)\\  
 m-XMT (True) & 0.677 (0.006) & 0.673 (0.007) & 0.665 (0.008) & 0.651 (0.010)\\  
 \hline 
 m-XMT (Main Effects) with $H_p^{(1)}(t)$ & 0.495 (0.021) & 0.581 (0.018) & 0.668 (0.012) & 0.730 (0.008)\\  
 \texttt{RFRE.PO} with $H_p^{(1)}(t)$ & 0.633 (0.007) & 0.653 (0.007) & 0.685 (0.006) & 0.712 (0.006)\\ 
 m-XMT (True) with $H_p^{(1)}(t)$ & 0.677 (0.007) & 0.674 (0.007) & 0.685 (0.007) & 0.711 (0.007)\\  
 \hline  
 m-XMT (Main Effects) with $H_f^{(1)}(t)$ & 0.522 (0.035) & 0.648 (0.017) & 0.687 (0.017) & 0.732 (0.010)\\  
 \texttt{RFRE.PO} with $H_f^{(1)}(t)$ & 0.636 (0.008) & 0.659 (0.008) & 0.688 (0.008) & 0.711 (0.007)\\ m-XMT (True) with $H_f^{(1)}(t)$ & 0.677 (0.006) & 0.678 (0.007) & 0.692 (0.010) & 0.711 (0.010)\\ 
 \hline 
 & & Light Censoring \\  \hline  m-XMT (Main Effects) & 0.533 (0.009) & 0.533 (0.009) & 0.534 (0.010) & 0.532 (0.011)\\ 
 \texttt{RFRE.PO} & 0.601 (0.015) & 0.593 (0.014) & 0.580 (0.017) & 0.562 (0.017)\\ 
 m-XMT (True) & 0.677 (0.007) & 0.672 (0.008) & 0.664 (0.009) & 0.650 (0.011)\\ 
 \hline  
 m-XMT (Main Effects) with $H_p^{(1)}(t)$ & 0.496 (0.023) & 0.574 (0.018) & 0.668 (0.012) & 0.724 (0.009)\\  
 \texttt{RFRE.PO} with $H_p^{(1)}(t)$ & 0.628 (0.008) & 0.649 (0.007) & 0.681 (0.007) & 0.708 (0.006)\\
 m-XMT (True) with $H_p^{(1)}(t)$ & 0.676 (0.006) & 0.674 (0.007) & 0.682 (0.008) & 0.705 (0.008)\\  
 \hline
 m-XMT (Main Effects) with $H_f^{(1)}(t)$ & 0.526 (0.032) & 0.643 (0.017) & 0.681 (0.019) & 0.725 (0.011)\\  
 \texttt{RFRE.PO} with $H_f^{(1)}(t)$ & 0.634 (0.008) & 0.656 (0.008) & 0.685 (0.008) & 0.708 (0.008)\\  
 m-XMT (True) with $H_f^{(1)}(t)$ & 0.677 (0.007) & 0.676 (0.008) & 0.689 (0.011) & 0.706 (0.011)\\ 
 \hline & & Moderate Censoring \\  
 \hline 
 m-XMT (Main Effects) & 0.534 (0.010) & 0.535 (0.010) & 0.535 (0.011) & 0.532 (0.012) \\  
 \texttt{RFRE.PO} & 0.597 (0.015) & 0.590 (0.015) & 0.577 (0.016) & 0.558 (0.017)\\ 
 m-XMT (True) & 0.677 (0.007) & 0.672 (0.008) & 0.664 (0.009) & 0.649 (0.012)\\ 
 \hline  
 m-XMT (Main Effects) with $H_p^{(1)}(t)$ & 0.495 (0.022) & 0.564 (0.018) & 0.661 (0.013) & 0.716 (0.011) \\ 
 \texttt{RFRE.PO} with $H_p^{(1)}(t)$ & 0.625 (0.009) & 0.646 (0.009) & 0.677 (0.008) & 0.704 (0.007) \\ 
 m-XMT (True) with $H_p^{(1)}(t)$ & 0.677 (0.007) & 0.674 (0.008) & 0.681 (0.008) & 0.700 (0.008) \\ 
 \hline
 m-XMT (Main Effects) with $H_f^{(1)}(t)$ & 0.529 (0.030) & 0.638 (0.017) & 0.676 (0.022) & 0.719 (0.013) \\ 
 \texttt{RFRE.PO} with $H_f^{(1)}(t)$ & 0.632 (0.009) & 0.654 (0.008) & 0.682 (0.009) & 0.705 (0.008)\\  
 m-XMT (True) with $H_f^{(1)}(t)$ & 0.676 (0.007) & 0.676 (0.008) & 0.688 (0.011) & 0.700 (0.010)\\  
 \hline  & & Heavy Censoring \\ 
 \hline
 m-XMT (Main Effects) & 0.535 (0.010) & 0.535 (0.011) & 0.535 (0.012) & 0.532 (0.012)\\ 
 \texttt{RFRE.PO} & 0.592 (0.015) & 0.585 (0.016) & 0.569 (0.017) & 0.551 (0.017)\\  
 m-XMT (True) & 0.677 (0.008) & 0.673 (0.008) & 0.664 (0.010) & 0.649 (0.012)\\  
 \hline 
 m-XMT (Main Effects) with $H_p^{(1)}(t)$ & 0.498 (0.022) & 0.557 (0.018) & 0.641 (0.015) & 0.707 (0.014) \\ 
 \texttt{RFRE.PO} with $H_p^{(1)}(t)$ & 0.621 (0.009) & 0.642 (0.008) & 0.672 (0.008) & 0.701 (0.007)\\ 
 m-XMT (True) with $H_p^{(1)}(t)$ & 0.677 (0.008) & 0.673 (0.008) & 0.678 (0.009) & 0.695 (0.009)\\  
 \hline 
 m-XMT (Main Effects) with $H_f^{(1)}(t)$ & 0.535 (0.031) & 0.635 (0.019) & 0.669 (0.025) & 0.709 (0.017)\\  
 \texttt{RFRE.PO} with $H_f^{(1)}(t)$ & 0.628 (0.010) & 0.652 (0.009) & 0.678 (0.010) & 0.701 (0.009)\\ 
 m-XMT (True) with $H_f^{(1)}(t)$ & 0.676 (0.008) & 0.676 (0.008) & 0.685 (0.011) & 0.695 (0.012)\\ 
 \hline  \end{tabular*}} \bigskip \end{table}

\begin{table}[!h]\footnotesize \caption{Simulated MSE listed as Mean (ESD) for varying prediction algorithms considered in Section \ref{s:SimStudy}, with censoring percentages (None, Light, Moderate, and Heavy), and gap-time correlations ($\rho = 0$, $0.3$, $0.6$, $0.9$). 500 replicates were performed with $n=500$ for each replicate.} \label{tab:simMSETable} {\begin{tabular*}{\textwidth}{@{}l@{\extracolsep{\fill}}c@{\extracolsep{\fill}}c@{\extracolsep{\fill}}c@{\extracolsep{\fill}}c@{}} 
\hline  
& $\rho = 0$ & $\rho = 0.3$ & $\rho = 0.6$ & $\rho = 0.9$ \\ 
\hline 
& & No Censoring \\  
\hline  
m-XMT (Main Effects) & 0.229 (0.003) & 0.212 (0.005) & 0.188 (0.006) & 0.159 (0.008) \\  
\texttt{RFRE.PO} & 0.213 (0.005) & 0.199 (0.005) & 0.180 (0.006) & 0.157 (0.008) \\ 
m-XMT (True) & 0.179 (0.003) & 0.164 (0.004) & 0.146 (0.005) & 0.127 (0.006) \\  
\hline  
m-XMT (Main Effects) with $H_p^{(1)}(t)$ & 0.235 (0.005) & 0.206 (0.006) & 0.168 (0.007) & 0.131 (0.008) \\ 
\texttt{RFRE.PO} with $H_p^{(1)}(t)$ & 0.202 (0.003) & 0.178 (0.005) & 0.147 (0.005) & 0.114 (0.005) \\  
m-XMT (True) with $H_p^{(1)}(t)$ & 0.179 (0.004) & 0.163 (0.005) & 0.140 (0.005) & 0.112 (0.006) \\ 

\hline 
m-XMT (Main Effects) with $H_f^{(1)}(t)$ & 0.231 (0.007) & 0.189 (0.009) & 0.157 (0.011) & 0.130 (0.009) \\
\texttt{RFRE.PO} with $H_f^{(1)}(t)$ & 0.200 (0.004) & 0.175 (0.004) & 0.146 (0.005) & 0.113 (0.005) \\ 
m-XMT (True) with $H_f^{(1)}(t)$ & 0.180 (0.004) & 0.162 (0.004) & 0.138 (0.005) & 0.112 (0.006) \\ 
\hline  & & Light Censoring \\  
\hline  m-XMT (Main Effects) & 0.233 (0.004) & 0.215 (0.005) & 0.190 (0.006) & 0.161 (0.008) \\  
\texttt{RFRE.PO} & 0.218 (0.005) & 0.203 (0.006) & 0.183 (0.007) & 0.160 (0.008) \\  
m-XMT (True) & 0.182 (0.004) & 0.167 (0.005) & 0.148 (0.006) & 0.129 (0.007) \\  
\hline 
m-XMT (Main Effects) with $H_p^{(1)}(t)$ & 0.238 (0.005) & 0.210 (0.006) & 0.172 (0.007) & 0.135 (0.008) \\ 
\texttt{RFRE.PO} with $H_p^{(1)}(t)$ & 0.207 (0.004) & 0.183 (0.005) & 0.152 (0.006) & 0.118 (0.006) \\ 
m-XMT (True) with $H_p^{(1)}(t)$ & 0.182 (0.004) & 0.166 (0.005) & 0.143 (0.006) & 0.115 (0.006) \\ 
\hline 
m-XMT (Main Effects) with $H_f^{(1)}(t)$ & 0.234 (0.006) & 0.195 (0.008) & 0.163 (0.011) & 0.134 (0.009) \\ 
\texttt{RFRE.PO} with $H_f^{(1)}(t)$ & 0.204 (0.004) & 0.180 (0.005) & 0.150 (0.006) & 0.118 (0.006) \\  
m-XMT (True) with $H_f^{(1)}(t)$ & 0.183 (0.004) & 0.166 (0.005) & 0.141 (0.006) & 0.116 (0.007) \\  
\hline 
& & Moderate Censoring \\
\hline 
m-XMT (Main Effects) & 0.236 (0.004) & 0.218 (0.005) & 0.193 (0.007) & 0.163 (0.008) \\ 
\texttt{RFRE.PO} & 0.222 (0.005) & 0.207 (0.006) & 0.186 (0.007) & 0.162 (0.008) \\  
m-XMT (True) & 0.185 (0.004) & 0.170 (0.005) & 0.151 (0.006) & 0.131 (0.007) \\  
\hline
m-XMT (Main Effects) with $H_p^{(1)}(t)$ & 0.242 (0.005) & 0.214 (0.006) & 0.172 (0.007) & 0.139 (0.008) \\  
\texttt{RFRE.PO} with $H_p^{(1)}(t)$ & 0.212 (0.004) & 0.188 (0.005) & 0.156 (0.006) & 0.122 (0.006) \\ 
m-XMT (True) with $H_p^{(1)}(t)$ & 0.186 (0.004) & 0.169 (0.005) & 0.146 (0.006) & 0.119 (0.006) \\ 
\hline 
m-XMT (Main Effects) with $H_f^{(1)}(t)$ & 0.237 (0.006) & 0.200 (0.008) & 0.169 (0.012) & 0.139 (0.009) \\  
\texttt{RFRE.PO} with $H_f^{(1)}(t)$ & 0.209 (0.004) & 0.184 (0.005) & 0.154 (0.006) & 0.122 (0.006) \\  
m-XMT (True) with $H_f^{(1)}(t)$ & 0.185 (0.004) & 0.168 (0.005) & 0.144 (0.006) & 0.119 (0.007) \\  
\hline 
& & Heavy Censoring \\ 
\hline  
m-XMT (Main Effects) & 0.240 (0.004) & 0.221 (0.005) & 0.197 (0.008) & 0.167 (0.009) \\ 
\texttt{RFRE.PO} & 0.228 (0.005) & 0.212 (0.006) & 0.191 (0.008) & 0.166 (0.009) \\  
m-XMT (True) & 0.190 (0.004) & 0.173 (0.005) & 0.155 (0.006) & 0.135 (0.007) \\ 
\hline  
m-XMT (Main Effects) with $H_p^{(1)}(t)$ & 0.246 (0.005) & 0.219 (0.006) & 0.182 (0.008) & 0.145 (0.009) \\  
\texttt{RFRE.PO} with $H_p^{(1)}(t)$ & 0.218 (0.004) & 0.194 (0.005) & 0.162 (0.006) & 0.127 (0.007) \\ 
m-XMT (True) with $H_p^{(1)}(t)$ & 0.190 (0.004) & 0.174 (0.005) & 0.151 (0.006) & 0.123 (0.007) \\  
\hline  
m-XMT (Main Effects) with $H_f^{(1)}(t)$ & 0.240 (0.006) & 0.205 (0.008) & 0.173 (0.011) & 0.145 (0.01) \\ 
\texttt{RFRE.PO} with $H_f^{(1)}(t)$ & 0.215 (0.004) & 0.189 (0.006) & 0.160 (0.007) & 0.128 (0.007) \\  
m-XMT (True) with $H_f^{(1)}(t)$ & 0.190 (0.004) & 0.172 (0.006) & 0.148 (0.007) & 0.124 (0.008) \\  
\hline  \end{tabular*}}  \bigskip  \end{table}

\clearpage
\begin{longtable}[ht] {c c  c c c c c   c c  c  c c}
\caption{Simulated proportion of times that the corresponding covariate appeared as statistically significant according to the Wald test for the m-XMT (Main Effects) method. History covariates considered were of the type $\tau^*$-restricted mean survival time in previous windows, where $\tau^*$ is $\frac{1}{12}$ years (described in further detail in Section \ref{sec:construct_history}). In the $H_p^{(1)}(t)$ case, results were from $m=10$ multiply imputed datasets combined via Rubin's rules. 500 replicates were performed with $n=500$ for each replicate. Covariates with signal ($Z_1, Z_2, Z_3, Z_4, Z_6$) are displayed first, and noise covariates are shown second ($Z_5$, $Z_7$). Within a censoring and correlation designation, quality of historical information included in the model increases, from top to bottom.} \label{tab:m-XMTSig}  \\  \multirow{2}{*}[-1em]{Censoring} & & \multicolumn{5}{c}{Signal} & \multicolumn{2}{c}{Noise} & \multicolumn{3}{c}{History Covariates Included} \\  \cmidrule(r){3-7} \cmidrule(r){8-9} \cmidrule(r){10-12} \cline{3-7} \cline{8-9}  \cline{10-12} & $\rho$ & $Z_1$ & $Z_2$ & $Z_3$ & $Z_4$ & $Z_6$ & $Z_5$ & $Z_7$ & $t$ & $H_p^{(1)}(t)$ & $H_f^{(1)}(t)$ \\  \hline  \endfirsthead  \caption[]{(continued)} \\  \multirow{2}{*}[-1em]{Censoring} & & \multicolumn{5}{c}{Signal} & \multicolumn{2}{c}{Noise} & \multicolumn{3}{c}{History Covariates Included} \\  \cmidrule(r){3-7}  \cmidrule(r){8-9}  \cmidrule(r){10-12}  \cline{3-7}  \cline{8-9}  \cline{10-12}  & $\rho$ & $Z_1$ & $Z_2$ & $Z_3$ & $Z_4$ & $Z_6$ & $Z_5$ & $Z_7$ & $t$ & $H_p^{(1)}(t)$ & $H_f^{(1)}(t)$ \\  \hline  \endhead  \hline  \multicolumn{12}{l}{\textit{Continued on next page}} \\  \hline  \endfoot  \endlastfoot 
None & 0.0 & 0.642 & 0.370 & 0.078 & 0.056 & 0.380 & 0.054 & 0.066 & 0.050 & -- & -- \\  
None & 0.0 & 0.641 & 0.293 & 0.050 & 0.077 & 0.378 & 0.054 & 0.085 & 0.046 & 0.402 & -- \\  
None & 0.0 & 0.646 & 0.372 & 0.078 & 0.060 & 0.380 & 0.060 & 0.066 & 0.046 & -- & 0.346 \\  
\hline 
None & 0.3 & 0.624 & 0.230 & 0.082 & 0.044 & 0.388 & 0.060 & 0.056 & 0.064 & -- & -- \\  
None & 0.3 & 0.635 & 0.232 & 0.076 & 0.057 & 0.374 & 0.052 & 0.052 & 0.071 & 0.431 & -- \\  
None & 0.3 & 0.528 & 0.222 & 0.074 & 0.050 & 0.248 & 0.054 & 0.040 & 0.358 & -- & 0.998 \\  
\hline 
None & 0.6 & 0.662 & 0.200 & 0.062 & 0.056 & 0.460 & 0.070 & 0.072 & 0.044 & -- & -- \\  
None & 0.6 & 0.495 & 0.168 & 0.060 & 0.071 & 0.304 & 0.043 & 0.092 & 0.522 & 1.000 & -- \\  
None & 0.6 & 0.470 & 0.172 & 0.062 & 0.054 & 0.190 & 0.092 & 0.072 & 0.672 & -- & 1.000 \\  
\hline  
None & 0.9 & 0.592 & 0.134 & 0.098 & 0.082 & 0.474 & 0.066 & 0.066 & 0.066 & -- & -- \\ 
None & 0.9 & 0.381 & 0.165 & 0.072 & 0.036 & 0.137 & 0.065 & 0.065 & 0.986 & 1.000 & -- \\  
None & 0.9 & 0.336 & 0.122 & 0.070 & 0.072 & 0.148 & 0.066 & 0.050 & 0.772 & -- & 1.000 \\  
\hline  
Low & 0.0 & 0.573 & 0.244 & 0.074 & 0.046 & 0.299 & 0.072 & 0.044 & 0.062 & -- & -- \\  
Low & 0.0 & 0.598 & 0.274 & 0.074 & 0.095 & 0.385 & 0.054 & 0.051 & 0.071 & 0.439 & -- \\  
Low & 0.0 & 0.581 & 0.246 & 0.082 & 0.044 & 0.303 & 0.066 & 0.046 & 0.066 & -- & 0.289 \\  
\hline  
Low & 0.3 & 0.592 & 0.206 & 0.072 & 0.060 & 0.398 & 0.054 & 0.040 & 0.052 & -- & -- \\ 
Low & 0.3 & 0.546 & 0.272 & 0.043 & 0.060 & 0.301 & 0.073 & 0.043 & 0.066 & 0.222 & -- \\ 
Low & 0.3 & 0.504 & 0.192 & 0.068 & 0.062 & 0.238 & 0.060 & 0.036 & 0.230 & -- & 0.998 \\  
\hline 
Low & 0.6 & 0.636 & 0.212 & 0.086 & 0.060 & 0.478 & 0.052 & 0.066 & 0.056 & -- & -- \\  
Low & 0.6 & 0.487 & 0.170 & 0.057 & 0.073 & 0.243 & 0.053 & 0.097 & 0.413 & 1.000 & -- \\  
Low & 0.6 & 0.474 & 0.192 & 0.072 & 0.074 & 0.226 & 0.068 & 0.070 & 0.528 & -- & 1.000 \\  
\hline 
Low & 0.9 & 0.582 & 0.168 & 0.096 & 0.094 & 0.500 & 0.054 & 0.064 & 0.052 & -- & -- \\  
Low & 0.9 & 0.351 & 0.120 & 0.060 & 0.043 & 0.187 & 0.060 & 0.060 & 0.967 & 1.000 & -- \\  
Low & 0.9 & 0.338 & 0.150 & 0.070 & 0.076 & 0.176 & 0.064 & 0.078 & 0.582 & -- & 1.000 \\  
\hline 
Medium & 0.0 & 0.618 & 0.250 & 0.068 & 0.060 & 0.358 & 0.094 & 0.088 & 0.080 & -- & -- \\ 
Medium & 0.0 & 0.580 & 0.279 & 0.061 & 0.065 & 0.370 & 0.059 & 0.057 & 0.055 & 0.475 & -- \\  
Medium & 0.0 & 0.622 & 0.248 & 0.066 & 0.064 & 0.370 & 0.094 & 0.084 & 0.084 & -- & 0.236 \\  
\hline 
Medium & 0.3 & 0.624 & 0.250 & 0.072 & 0.074 & 0.372 & 0.066 & 0.060 & 0.066 & -- & -- \\  
Medium & 0.3 & 0.599 & 0.246 & 0.068 & 0.054 & 0.393 & 0.086 & 0.052 & 0.072 & 0.132 & -- \\ 
Medium & 0.3 & 0.508 & 0.230 & 0.080 & 0.080 & 0.234 & 0.070 & 0.064 & 0.172 & -- & 0.994 \\ 
\hline 
Medium & 0.6 & 0.620 & 0.160 & 0.078 & 0.074 & 0.464 & 0.060 & 0.058 & 0.060 & -- & -- \\ 
Medium & 0.6 & 0.485 & 0.190 & 0.080 & 0.052 & 0.251 & 0.064 & 0.052 & 0.289 & 1.000 & -- \\ 
Medium & 0.6 & 0.442 & 0.166 & 0.074 & 0.070 & 0.236 & 0.072 & 0.054 & 0.410 & -- & 1.000 \\ 
\hline  
Medium & 0.9 & 0.582 & 0.182 & 0.076 & 0.062 & 0.488 & 0.042 & 0.062 & 0.052 & -- & -- \\  
Medium & 0.9 & 0.412 & 0.151 & 0.050 & 0.072 & 0.207 & 0.062 & 0.054 & 0.922 & 1.000 & -- \\  
Medium & 0.9 & 0.376 & 0.176 & 0.068 & 0.064 & 0.190 & 0.042 & 0.064 & 0.466 & -- & 1.000 \\  
\hline  
High & 0.0 & 0.580 & 0.254 & 0.060 & 0.066 & 0.322 & 0.048 & 0.054 & 0.058 & -- & -- \\  
High & 0.0 & 0.584 & 0.263 & 0.066 & 0.078 & 0.400 & 0.056 & 0.068 & 0.066 & 0.436& -- \\  
High & 0.0 & 0.584 & 0.246 & 0.062 & 0.064 & 0.322 & 0.048 & 0.058 & 0.056 & -- & 0.244\\  
\hline 
High & 0.3 & 0.562 & 0.260 & 0.080 & 0.078 & 0.358 & 0.064 & 0.076 & 0.084 & --& -- \\  
High & 0.3 & 0.583 & 0.228 & 0.070 & 0.060 & 0.315 & 0.040 & 0.114 & 0.068 & 0.044 & -- \\  
High & 0.3 & 0.462 & 0.228 & 0.078 & 0.078 & 0.238 & 0.058 & 0.078 & 0.146 & -- & 0.988 \\  
\hline 
High & 0.6 & 0.570 & 0.198 & 0.074 & 0.076 & 0.424 & 0.088 & 0.048 & 0.068 & -- & -- \\  
High & 0.6 & 0.536 & 0.216 & 0.074 & 0.064 & 0.304 & 0.056 & 0.058 & 0.198 & 0.984 & -- \\  
High & 0.6 & 0.428 & 0.174 & 0.082 & 0.074 & 0.208 & 0.064 & 0.044 & 0.302 & -- & 1.000 \\  
\hline  
High & 0.9 & 0.570 & 0.160 & 0.088 & 0.076 & 0.460 & 0.064 & 0.066 & 0.068 & -- & -- \\  
High & 0.9 & 0.433 & 0.122 & 0.064 & 0.068 & 0.230 & 0.066 & 0.034 & 0.770 & 1.000 & -- \\  
High & 0.9 & 0.390 & 0.154 & 0.076 & 0.082 & 0.226 & 0.058 & 0.070 & 0.276 & -- & 1.000 \\ 
\end{longtable}

\clearpage
\begin{longtable}[h]{cc ccccc ccc ccc}  \caption{Simulated proportion of times that the corresponding covariate appeared as statistically significant according to the random forest permutation test. 500 replicates were performed, with $n=500$ subjects in each replicate. History covariates considered were of the type $\tau^*$-restricted mean survival time in previous windows, where $\tau^*$ is $\frac{1}{12}$ years (described in further detail in Section \ref{sec:construct_history}). In the $H_p^{(1)}(t)$ case, results were from $m=10$ multiply imputed datasets, combined via Rubin's rules. Covariates with signal ($Z_1, Z_2, Z_3, Z_4, $ and $Z_6$) are displayed first, and noise covariates are shown second ($Z_5$, $Z_7$). Within a censoring and correlation designation, quality of historical information included in the model increases, from top to bottom.}\label{tab:checkPower} \\  \multirow{2}{*}[-1em]{Censoring} & & \multicolumn{5}{c}{Signal} & \multicolumn{2}{c}{Noise} & \multicolumn{3}{c}{History Covariates Included} \\  \cmidrule(r){3-7} \cmidrule(r){8-9} \cmidrule(r){10-12} \cline{3-7} \cline{8-9}  \cline{10-12} & $\rho$ & $Z_1$ & $Z_2$ & $Z_3$ & $Z_4$ & $Z_6$ & $Z_5$ & $Z_7$ & $t$ & $H_p^{(1)}(t)$ & $H_f^{(1)}(t)$ \\  \hline  \endfirsthead  \caption[]{(continued)} \\  \multirow{2}{*}[-1em]{Censoring} & & \multicolumn{5}{c}{Signal} & \multicolumn{2}{c}{Noise} & \multicolumn{3}{c}{History Covariates Included} \\  \cmidrule(r){3-7}  \cmidrule(r){8-9}  \cmidrule(r){10-12}  \cline{3-7}  \cline{8-9}  \cline{10-12}  & $\rho$ & $Z_1$ & $Z_2$ & $Z_3$ & $Z_4$ & $Z_6$ & $Z_5$ & $Z_7$ & $t$ & $H_p^{(1)}(t)$ & $H_f^{(1)}(t)$ \\  \hline  \endhead  \hline  \multicolumn{12}{l}{\textit{Continued on next page}} \\  \hline  \endfoot  \endlastfoot 

None & 0.00 & 1.000 & 1.000 & 0.954 & 0.066 & 1.000 & 0.050 & 0.050 & 0.062 & -- & -- \\ 
None & 0.0 & 1.000 & 0.552 & 0.008 & 0.000 & 1.000 & 0.000 & 0.000 & 0.538 & 1.000 & -- \\
None & 0.00 & 1.000 & 0.988 & 0.866 & 0.050 & 1.000 & 0.038 & 0.036 & 0.992 & -- & 1.000 \\ 
\hline 
None & 0.3 & 1.000 & 0.994 & 0.838 & 0.052 & 1.000 & 0.044 & 0.034 & 0.046 & -- & -- \\ 
None & 0.3 & 1.000 & 0.332 & 0.006 & 0.002 & 1.000 & 0.000 & 0.002 & 0.994 & 1.000 & -- \\ 
None & 0.3 & 1.000 & 0.942 & 0.578 & 0.032 & 1.000 & 0.044 & 0.038 & 0.996 & -- & 1.00\\ 
\hline 
None & 0.6 & 1.000 & 0.954 & 0.610 & 0.048 & 1.000 & 0.038 & 0.042 & 0.046 & -- & -- \\ 
None & 0.6 & 1.000 & 0.223 & 0.010 & 0.026 & 1.000 & 0.006 & 0.000 & 1.000 & 1.000 & -- \\ 
None & 0.6 & 1.000 & 0.808 & 0.356 & 0.032 & 1.000 & 0.036 & 0.036 & 1.000  & -- & 1.00\\ 
\hline 
None & 0.9 & 1.000 & 0.792 & 0.334 & 0.032 & 1.000 & 0.032 & 0.034 & 0.046 & -- & -- \\ 
None & 0.9 & 1.000 & 0.123 & 0.012 & 0.127 & 0.998 & 0.002 & 0.006 & 1.000 & 1.000 & -- \\
None & 0.9 & 1.000 & 0.488 & 0.136 & 0.034 & 1.000 & 0.034 & 0.026 & 1.000 & -- & 1.00\\ 
\hline 
Low & 0.0 & 1.000 & 1.000 & 0.890 & 0.044 & 1.000 & 0.020 & 0.030 & 0.022 & -- & -- \\ 
Low & 0.0 & 1.000 & 0.413 & 0.004 & 0.000 & 1.000 & 0.000 & 0.000 & 0.451 & 1.000 & -- \\ 
Low & 0.0 & 1.000 & 0.970 & 0.739 & 0.038 & 1.000 & 0.014 & 0.026 & 0.824 & -- & 1.00\\ 
\hline 
Low & 0.3 & 1.000 & 0.982 & 0.780 & 0.032 & 1.000 & 0.038 & 0.044 & 0.038 & -- & -- \\ 
Low & 0.3 & 1.000 & 0.280 & 0.008 & 0.002 & 1.000 & 0.000 & 0.000 & 0.976 & 1.000 & -- \\
Low & 0.3 & 1.000 & 0.884 & 0.476 & 0.028 & 1.000 & 0.028 & 0.030 & 0.840 & -- & 1.00\\ 
\hline 
Low & 0.6 & 1.000 & 0.938 & 0.506 & 0.018 & 1.000 & 0.016 & 0.040 & 0.042 & -- & -- \\ 
Low & 0.6 & 1.000 & 0.172 & 0.010 & 0.024 & 1.000 & 0.002 & 0.002 & 1.000 & 1.000 & -- \\ 
Low & 0.6 & 1.000 & 0.712 & 0.260 & 0.022 & 1.000 & 0.016 & 0.040 & 0.948 & -- & 1.00\\ 
\hline
Low & 0.9 & 1.000 & 0.716 & 0.264 & 0.052 & 1.000 & 0.016 & 0.034 & 0.048 & -- & -- \\
Low & 0.9 & 0.996 & 0.118 & 0.006 & 0.104 & 1.000 & 0.008 & 0.024 & 1.000 & 1.000 & -- \\ 
Low & 0.9 & 1.000 & 0.424 & 0.122 & 0.022 & 1.000 & 0.024 & 0.018 & 0.996 & -- & 1.000 \\ 
\hline 
Medium & 0.0 & 1.000 & 0.998 & 0.832 & 0.032 & 1.000 & 0.032 & 0.030 & 0.042 & -- & -- \\ 
Medium & 0.0 & 1.000 & 0.303 & 0.004 & 0.000 & 1.000 & 0.000 & 0.000 & 0.403 & 1.000 & -- \\ 
Medium & 0.0 & 1.000 & 0.926 & 0.618 & 0.034 & 1.000 & 0.042 & 0.026 & 0.696 & -- & 1.00\\ 
\hline 
Medium & 0.3 & 1.000 & 0.982 & 0.694 & 0.038 & 1.000 & 0.026 & 0.030 & 0.070 & -- & -- \\
Medium & 0.3 & 1.000 & 0.222 & 0.010 & 0.006 & 1.000 & 0.000 & 0.000 & 0.967 & 1.000 & -- \\ 
Medium & 0.3 & 1.000 & 0.818 & 0.406 & 0.032 & 1.000 & 0.022 & 0.012 & 0.638 & -- & 1.00\\ 
\hline 
Medium & 0.6 & 1.000 & 0.898 & 0.448 & 0.036 & 1.000 & 0.020 & 0.022 & 0.064 & -- & -- \\ 
Medium & 0.6 & 1.000 & 0.161 & 0.010 & 0.024 & 1.000 & 0.000 & 0.000 & 1.000 & 1.000 & -- \\ 
Medium & 0.6 & 1.000 & 0.622 & 0.196 & 0.020 & 1.000 & 0.012 & 0.020 & 0.824 & -- & 1.00\\ 
\hline 
Medium & 0.9 & 1.000 & 0.688 & 0.258 & 0.028 & 1.000 & 0.016 & 0.016 & 0.064 & -- & -- \\ 
Medium & 0.9 & 1.000 & 0.113 & 0.014 & 0.138 & 1.000 & 0.006 & 0.023 & 1.000 & 1.000 & -- \\
Medium & 0.9 & 1.000 & 0.440 & 0.112 & 0.034 & 1.000 & 0.020 & 0.018 & 0.926 & -- & 1.00\\ 
\hline 
High & 0.0 & 1.000 & 0.966 & 0.688 & 0.036 & 1.000 & 0.024 & 0.020 & 0.060 & -- & -- \\ 
High & 0.0 & 1.000 & 0.211 & 0.000 & 0.000 & 1.000 & 0.000 & 0.000 & 0.310 & 1.000 & -- \\ 
High & 0.0 & 1.000 & 0.836 & 0.490 & 0.038 & 1.000 & 0.022 & 0.022 & 0.582 & -- & 1.00\\ 
\hline 
High & 0.3 & 1.000 & 0.946 & 0.506 & 0.028 & 1.000 & 0.014 & 0.016 & 0.098 & -- & -- \\ 
High & 0.3 & 1.000 & 0.192 & 0.004 & 0.002 & 1.000 & 0.000 & 0.000 & 0.953 & 1.000 & -- \\
High & 0.3 & 1.000 & 0.706 & 0.268 & 0.018 & 1.000 & 0.012 & 0.016 & 0.590 & -- & 1.000 \\ 
\hline
High & 0.6 & 1.000 & 0.812 & 0.368 & 0.014 & 1.000 & 0.022 & 0.026 & 0.094 & -- & -- \\ 
High & 0.6 & 1.000 & 0.126 & 0.010 & 0.024 & 1.000 & 0.002 & 0.006 & 1.000 & 1.000 & -- \\ 
High & 0.6 & 1.000 & 0.524 & 0.138 & 0.020 & 1.000 & 0.018 & 0.026 & 0.716 & -- & 1.000 \\ 
\hline
High & 0.9 & 1.000 & 0.602 & 0.226 & 0.016 & 1.000 & 0.016 & 0.014 & 0.086 & -- & -- \\
High & 0.9 & 0.998 & 0.106 & 0.033 & 0.110 & 0.994 & 0.004 & 0.018 & 1.000 & 1.000 & -- \\
High & 0.9 & 1.000 & 0.314 & 0.094 & 0.018 & 1.000 & 0.022 & 0.012 & 0.818 & -- & 1.000 \\  
\hline 
\end{longtable}

\clearpage
\section{Tables and Figures for the Analysis of the Azithromycin for the Prevention of COPD Cohort}

\begin{figure}[h]
    \centering \includegraphics[width=\textwidth]{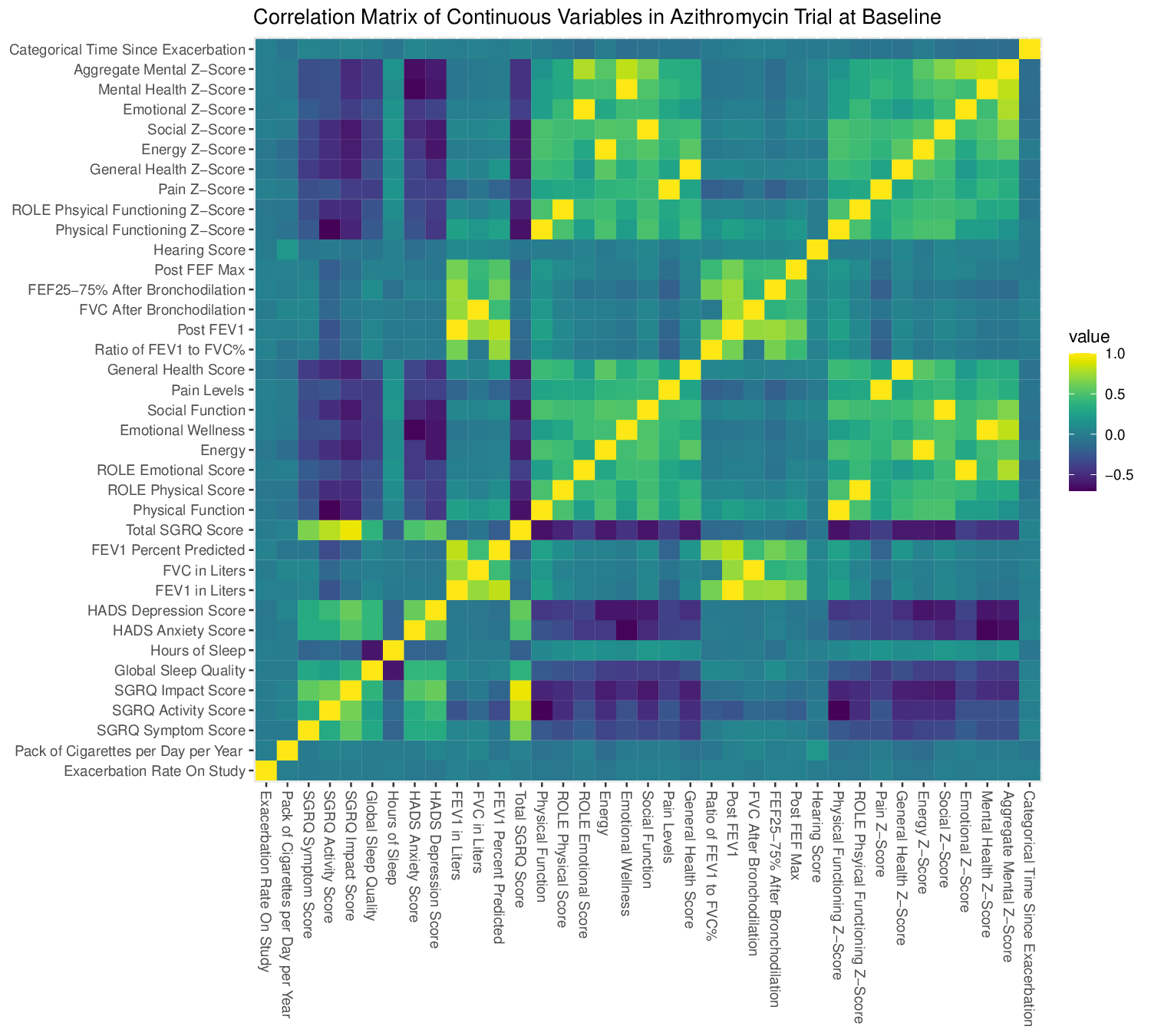}
    \caption{A correlation heat map of continuous and ordinal variables in the Azithromycin for the Prevention of COPD cohort, measured at baseline. Variables derived from each other have instances of particularly high positive or negative correlation. These correlation patterns make random forest a more attractive modeling alternative compared to semi-parametric and parametric methods.}\label{fig:corrHeatMap}
\end{figure}

\begin{table}[h]  
 \caption{m-XMT parameter estimates, based on covariates selected by \citet{Xia:StatMed:2019:regression}, applied to censored longitudinal data with $\tau = 180$, $a = 30$.} \label{tab:summer-estimates}
 {\begin{tabular*}{\columnwidth}{@{}l@{\extracolsep{\fill}}c@{\extracolsep{\fill}}c@{\extracolsep{\fill}}r@{}}
 \hline
  Covariate & $e^{\hat{\beta} \dag}$ & 95\% CI &  p\\
  \hline\hline
  Azithromycin (versus placebo) & 1.43 & (1.14, 1.79) & $<$\textbf{0.01} \\ 
  Smoker at baseline (versus non-smoker) & 1.07 & (0.80, 1.43) & 0.66 \\ 
  FEV$_1$ in liters at baseline & 1.32 & (1.05, 1.66) & \textbf{0.02} \\ 
  Male gender (versus female) & 1.36 & (1.07, 1.74) & \textbf{0.01} \\ 
  Age in decades & 1.14 & (0.99, 1.32) & 0.07 \\ 
 \hline
 \end{tabular*}}
 \small{$^\dag$ $e^{\hat{\beta}}$ is the estimated odds ratio for being event-free during a 180-day follow-up period corresponding to a 1-unit increase in the (continuous) predictor, or the presence of a binary covariate, adjusted for other predictors.

FEV$_1$ is forced expiratory volume in one second, a clinical measure of lung function.}
\bigskip
\end{table}

\begin{table}[h] 
 \caption{Parameter estimates from the m-XMT (Forward Selection) model based on the first of $m = 10$ imputed datasets. For the analysis, we created three history variables, $H_f(t), H_{p}^{(2)}(t)$ and $H_{p}^{(3)}(t)$. $H_f(t)$ was an indicator if the subject had experienced a severe exacerbation on study time prior to time $t$. $H_{p}^{(2)}(t)$ was assumed to take a categorical form based on whether the participant had an exacerbation in the past zero to 31 days ($H_{p}^{(2)}(t) =4$), 31-92 days ($H_{p}^{(2)}(t) =3$), 93-182 days ($H_{p}^{(2)}(t) =2$), 183 days to 365 days ($H_{p}^{(2)}(t) =1$), or finally, if that time was more than 365 days or never ($H_{p}^{(2)}(t) =0$). $H_{p}^{(3)}(t)$ is the estimated 30-day event rate using all follow-up prior to time $t$. A very recent exacerbation, bronchitis, gender, treatment, the ratio of FEV to FVC \%predicted, SGRQ symptoms, oxygen use, recent hospitalization, rate of exacerbations, and use of ICS medication  remain significant across the $m=10$ imputed datasets. 
 } \label{table:wald-estimates}

 {\begin{tabular*}{\columnwidth}{@{}l@{\extracolsep{\fill}}c@{\extracolsep{\fill}}c@{\extracolsep{\fill}}r@{\extracolsep{\fill}}r@{}}
 \hline
  Covariate & $e^{\hat{\beta}}$ $^\dag$ & 95\% CI & p\\
  \hline\hline
   Categorical $H_{p,i}^{(2)}(t)= 0$ (versus 1, 2, 3 or 4) & 2.32 & (2.01, 2.68)  & $<$\textbf{0.01} \\
  Hospitalization in the year before $t=0$ (versus not) & 0.70 & (0.53, 0.92) & \textbf{0.01} \\ 
  Corticosteroid use in the year before $t=0$ (versus no use)& 0.61 & (0.40, 0.94) & \textbf{0.03} \\ 
  $H_{f,i}(t)= 1$ (versus $H_{f, i}(t) = 0$) & 1.81 & (1.23, 2.68) & \textbf{0.01} \\
  Time $t$ (in months) & 0.96 & (0.93, 1.00) & \textbf{0.02} \\ 
  Male gender (versus female gender) & 1.69 & (1.28, 2.23) & $<$\textbf{0.01} \\ 
  Race = Black (versus non-Black) & 0.57 & (0.37, 0.89) & \textbf{0.01} \\ 
  Sudden feelings of panic at time $t$ & 0.84 & (0.74, 0.95) & \textbf{0.01} \\
  \hspace{1cm} (versus no such feelings) \\
  Sudden feelings of \say{butterfly} fright at time $t$ & 0.89 & (0.78, 1.01) & 0.08\\
  \hspace{1cm} (versus no such feelings)\\
  Seen a mental health provider at time $t$ & 0.75 & (0.61, 0.93) & \textbf{0.01} \\ 
  \hspace{1cm} (versus has not seen a provider)\\
  Has pain at time $t$ (versus no pain) & 0.92 & (0.86, 0.98) & \textbf{0.03} \\
  St. George's Respiratory Questionnaire symptoms score at time $t$ & 0.99 & (0.99, 1.00) & \textbf{0.01} \\
  Antiarrhythmic use at time $t$ (versus no use) & 0.07 & (0.00, 6.56) & 0.25 \\ 
  FEV$_1$/FVC \% predicted at time $t$ & 1.01 & (1.00, 1.02) & $<$\textbf{0.01} \\ 
  Azithromycin treatment group (versus placebo) & 0.65 & (0.49, 0.85) & $<$\textbf{0.01} \\  
   Inhaled corticosteroid or long-acting beta-agonist use at time $t$ & 1.28 & (1.03, 1.60) & \textbf{0.02} \\ 
   Other baseline medications & 0.15 & (0.03, 0.81) & \textbf{0.02} \\ 
  \hspace{1cm} (versus no miscellaneous medications)\\ 
    Supplemental oxygen use at time $t$ & 0.55 & (0.41, 0.73) & $<$\textbf{0.01}\\ 
  \hspace{1cm} (versus no supplemental oxygen) \\
  Mixed beta-blocker use at time $t$ & 3.79 & (1.08, 13.24) & \textbf{0.04} \\ 
  \hspace{1cm}(versus selective or no beta-blocker use)\\
 Bronchiectasis diagnosis at time $t$ (versus no such diagnosis)& 0.64 & (0.48, 0.85) & $<$\textbf{0.01}\\ 
 \hline
 \end{tabular*}}
 \small{$^\dag$: $e^{\hat{\beta}}$ is the estimated odds ratio for being event-free during a 180-day follow-up period corresponding to a 1-unit increase in the (continuous) predictor, or the presence of a binary covariate, adjusted for other predictors.

 FEV$_1$ is forced expiratory volume in one second and FVC is forced vital capacity, clinical measures of lung function.

 SGRQ is the Saint George's Respiratory Questionnaire, a patient reported outcome measuring quality of life.

 }
\bigskip
\end{table}

\begin{table}[h] 
 \caption{P-values from the \texttt{\texttt{RFRE.PO}} permutation test applied to the out-of-bag-samples combined across $m = 10$ multiply imputed datasets. Included predictors had a statistically significant permutation test z-score for at least one of the multiply imputed training datasets. Additional predictors that did not achieve statistical significance via the permutation test, but were used in the \texttt{\texttt{RFRE.PO}} algorithm in some form, are not shown. For analysis, we created three history variables, $H_f(t), H_{p}^{(2)}(t)$ and $H_{p}^{(3)}(t)$. $H_f(t)$ was an indicator if the subject had experienced a severe exacerbation on study time prior to time $t$. $H_{p}^{(2)}(t)$ was assumed to take a categorical form based on whether the participant had an exacerbation in the past zero to 31 days ($H_{p}^{(2)}(t) =4$), 31-92 days ($H_{p}^{(2)}(t) =3$), 93-182 days ($H_{p}^{(2)}(t) =2$), 183 days to 365 days ($H_{p}^{(2)}(t) =1$), or finally, if that time was more than 365 days or never ($H_{p}^{(2)}(t) =0$). $H_{p}^{(3)}(t)$ is the estimated 30-day event rate using all follow-up prior to time $t$.}  \label{table:perm-tests}
 {\begin{tabular*}{\columnwidth}{@{}l@{\extracolsep{\fill}}r@{\extracolsep{\fill}}r@{}}
 \hline
  Covariate &p\\
  \hline\hline
  Categorical $H_{p,i}^{(2)}(t)= 0$ & 0.06\\
  \hspace{1cm} (versus Categorical $H_{p,i}^{(2)}(t)= 1, 2, 3, 4$) &\\
  Categorical $H_{p,i}^{(2)}(t)= 1$ & $<$\textbf{0.04}\\
  \hspace{1cm} (versus Categorical $H_{p,i}^{(2)}(t)=0, 2, 3, 4$) &\\
  Categorical $H_{p,i}^{(2)}(t)= 4$ & $<$\textbf{0.01}\\
  \hspace{1cm} (versus Categorical $H_{p,i}^{(2)}(t)=0, 1, 2, 3$) &\\
Categorical $H_{p, i}^{(2)}(t)$ & $<$\textbf{0.01}\\
$H_{p,i}^{(3)}(t)$ & $<$\textbf{0.01}\\
$H_{f,i}(t)$ & $<$\textbf{0.01} \\
Respiratory Hospitalization in the Year before Study Start & \textbf{0.01}\\
Corticosteroid Use in the Year before Study Start & $<$\textbf{0.01}\\
Gender & 0.99\\
Smoking at Baseline & 0.98\\
General Health Score at Time $t$ & $<$\textbf{0.01}\\
Standardized General Health Z-Score at Time $t$ & $<$\textbf{0.01}\\
Pain Score at Time $t$ & \textbf{0.03}\\
Standardized Pain Score at Time $t$ & \textbf{0.05}\\
SGRQ Symptom Score at Time $t$ & $<$\textbf{0.01}\\
Seen a Mental Health Professional for Anxiety/Depression at Time $t$ & 1.00\\
Anti-Coagulant Use at Time t & 1.00\\
Inhaled Corticosteroids, Long-Acting Muscarinic, & 0.99\\
\hspace{1cm} and Long-Acting Bronchodilator Agent at Time $t$ \\
Long-Acting Muscarinic Use at Time $t$ & 1.00\\
Treatment Group & \textbf{0.02}\\
Leukocytosis at Time $t$ & \textbf{0.02}\\
Oxygen Use at Time $t$ & 1.00\\
Bronchiectasis Diagnosis at Time $t$ & $<$\textbf{0.01}\\
Phlegm at Time $t$ & 1.00\\
 \hline
 \end{tabular*}}
  \small{
 FEV$_1$ is forced expiratory volume in one second and FVC is forced vital capacity in liters, clinical measures of lung function.

 SGRQ is the Saint George's Respiratory Questionnaire, a patient reported outcome measuring quality of life.

 }
\bigskip
\end{table}

\begin{table}[h]
 \caption{Comparison of the C-statistics and mean-squared errors (MSE) of models fit to the validation set of the Azithromycin in the Prevention of COPD cohort. The Wald forward selection model and the model corresponding to clinical input from \citet{Xia:StatMed:2019:regression} are fit using the modified XMT approach. The forward selection model includes predictors through automated forward selection requiring $p <0.05$ for entry into the model. Missing values in the forward selection model and \texttt{RFRE.PO} for time since most recent exacerbation are multiply imputed using the coding scheme described in Section \ref{s:data-app}. C-statistic and mean-squared error estimates and standard errors for all methods and models are averaged across results from multiply imputed data with Rubin's rule applied to bootstrapped ($b=100$) standard errors.}\label{table:data-res}
 \centering
 {\begin{tabularx}{0.85\textwidth}{@{}l@{\extracolsep{\fill}}cc@{\extracolsep{\fill}}cc@{}}
 \hline
 \multirow{2}{*}{Model Name} & \multicolumn{2}{c}{C-Statistic} & \multicolumn{2}{c}{MSE}\\
 \cmidrule(r){2-3} \cmidrule(l){4-5}
                             & {\small Mean} & {\small SD} & {\small Mean} & {\small SD}\\
 \hline\hline
 \texttt{RFRE.PO} & $0.613$ & $0.009$ & $0.242$ & $0.003$ \\
 m-XMT (Clinical Input) & $0.565$ &  $0.009$ & $0.252$ & $0.002$ \\
 m-XMT (Forward Selection) \hspace{5mm} & $0.570$ & $0.009$ & $0.264$ & $0.004$ \\
 \hline
 \end{tabularx}}
\bigskip
\end{table}

\clearpage
\begin{figure}
    \centering \includegraphics[width=\textwidth]{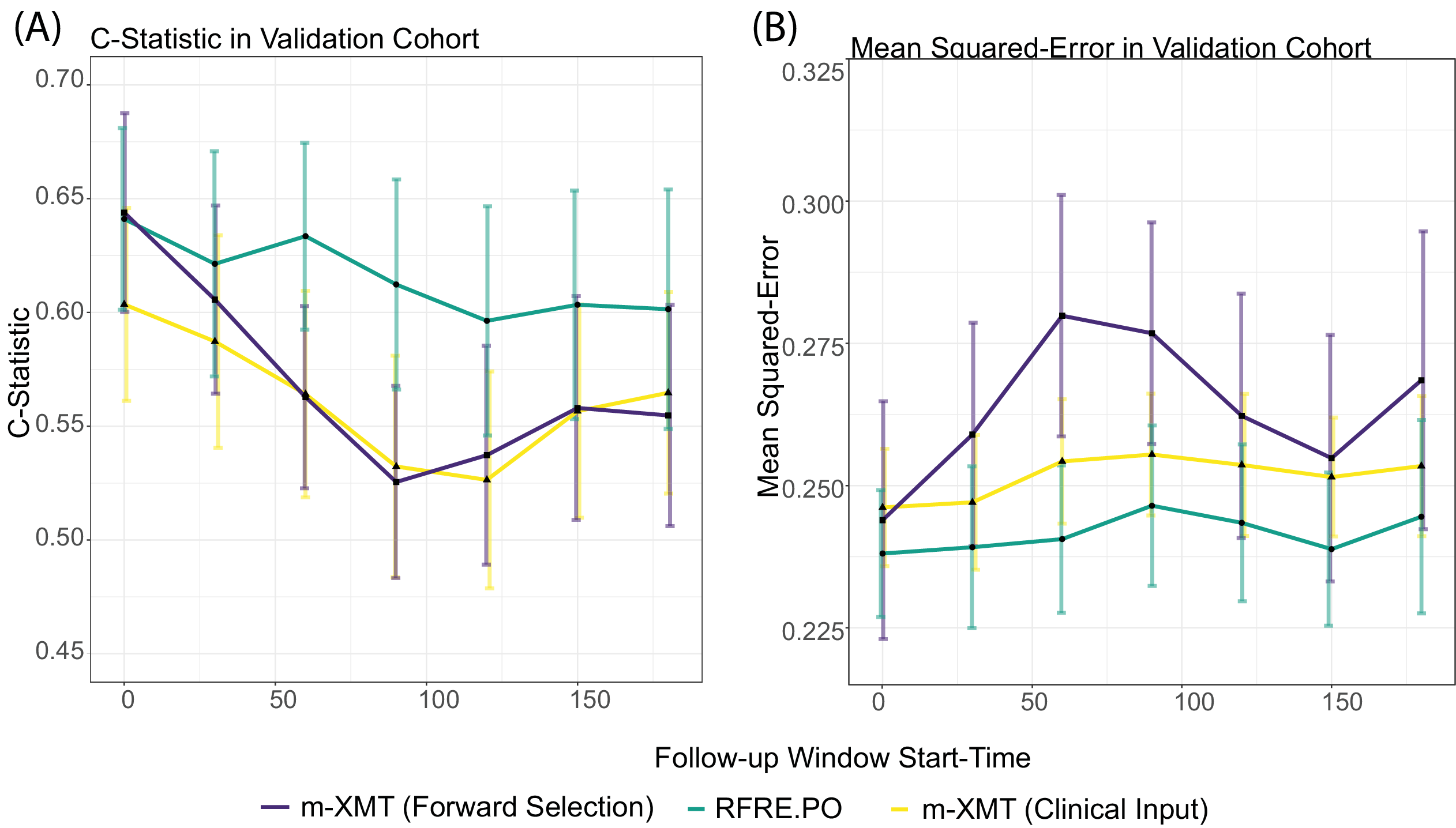}
    \caption{A panel graphic displaying time-dependent C-statistics (Panel A) and MSEs (Panel B) in the validation dataset of the Azithromycin for the Prevention of COPD study, calculated at each follow-up window $t\in \{0, 30, 60, 90, 120, 150, 180\}$ for $\tau=180$; 95\% confidence intervals also displayed. 
    Compared to the other algorithms, \texttt{RFRE.PO} has fairly stable and attractive performance metrics across the various follow-up windows, likely because interactions with the various predictors over time are naturally incorporated if useful for prediction. 
    }
    \label{fig:TimeVaryingCMSE}
\end{figure}




\clearpage


\bibliography{rfrepo}
\bibliographystyle{apalike}
